\documentclass[twocolumn]{aastex62}
\usepackage{graphicx,float,subfigure} 
\usepackage{float}
\usepackage[T1]{fontenc}
\usepackage{ae}
\usepackage{aecompl}
\usepackage{amsmath, amssymb, amsfonts, braket, gensymb} 
\usepackage{newtxtext,newtxmath} 
\usepackage{enumitem}
\def\url#1{\expandafter\string\csname #1\endcsname}

\received{June 19, 2018}
\revised{??? ??, 2018}
\accepted{\today}
\submitjournal{ApJ}

\shorttitle{The Mira-based distance to the Galactic centre}
\shortauthors{Qin et al.}

\begin{document}

\title{The Mira-based distance to the Galactic centre}
\correspondingauthor{Wenzer Qin}
\email{wqin2@jhu.edu, wenzer.qin@gmail.com}

\correspondingauthor{David M. Nataf}
\email{dnataf1@jhu.edu, david.nataf@gmail.com}

\author{Wenzer Qin}
\affil{Center for Astrophysical Sciences and Department of Physics and Astronomy, \\
The Johns Hopkins University, \\
Baltimore, MD 21218}
\author{David M. Nataf}
\affiliation{Center for Astrophysical Sciences and Department of Physics and Astronomy, \\
The Johns Hopkins University, \\
Baltimore, MD 21218}
\author{Nadia Zakamska}
\affiliation{Center for Astrophysical Sciences and Department of Physics and Astronomy, \\
The Johns Hopkins University, \\
Baltimore, MD 21218}
\author{Peter R. Wood}
\affiliation{Research School of Astronomy and Astrophysics, \\
Australian National University, Canberra, \\ 
ACT 2611, Australia \\}
\author{Luca Casagrande}
\affiliation{Research School of Astronomy and Astrophysics, \\
Australian National University, Canberra, \\ 
ACT 2611, Australia \\}

	
	\begin{abstract}
		Mira variables are useful distance indicators, due to their high luminosities and well-defined period-luminosity relation. We select 1863 Miras from SAAO and MACHO observations to examine their use as distance estimators in the Milky Way. We measure a distance to the Galactic centre of $R_0 = 7.9 \pm 0.3$ kpc, which is in good agreement with other literature values. The uncertainty has two components of $\sim$0.2 kpc each: the first is from our analysis and predominantly due to interstellar extinction, the second is due to zero-point uncertainties extrinsic to our investigation, such as the distance to the Large Magellanic Cloud (LMC). In an attempt to improve existing period-luminosity calibrations, we use theoretical models of Miras to determine the dependence of the period-luminosity  relation on age, metallicity, and helium abundance, under the assumption that Miras trace the bulk stellar population. We find that at a fixed period of $\log P = 2.4$, changes in the predicted $K_s$ magnitudes can be approximated by $\Delta M_{Ks} \approx -0.109(\Delta \rm{[Fe/H]}) + 0.033( {\Delta}t/\rm{Gyr}) + 0.021 ({\Delta}Y/0.01)$, and these coefficients are nearly independent of period. The expected overestimate in the Galactic centre distance from using an LMC-calibrated relation  is $\sim$0.3 kpc. This prediction is not validated by our analysis; a few possible reasons are discussed.  We separately show that while the predicted color-color diagrams of solar-neighbourhood Miras work well in the near-infrared, though there are offsets from the model predictions in the optical and mid-infrared.
	\end{abstract}
	

	\section{Introduction} \label{sec:Introduction}
	
	Measurements of Hubble's constant, i.e. the current expansion rate of the universe, are of great interest in modern astrophysics, since its value is a fundamental parameter of $\Lambda$-CDM cosmology. \cite{2012ApJ...758...24F} and \citet{2018ApJ...855..136R} have respectively measured Hubble's constant in the local universe to 3.5\% and 2.3\% uncertainty. These values are now in tension with other measurements, such as those determined from the cosmic microwave background \citep{2017arXiv170706547A,2018arXiv180706209P}. This tension might be due to ground-breaking new physics, so to study the discrepancy it is critical to probe and extend the local distance ladder by independent means.
	
	Mira variables provide a plausible extension to the extragalactic distance scale. They are bright in the infrared for both intermediate-age and old stellar populations, fairly numerous, and have a well-defined period-luminosity relation \citep{1989MNRAS.241..375F,1990AJ.....99..784H,2008MNRAS.386..313W}. Thus, one can envisage future catalogues of Mira variables toward great distances produced from \textit{James Webb Space Telescope (JWST)} photometry. 	Miras are pulsating variable stars that lie in the late evolutionary stages of the asymptotic giant branch (AGB). They are characterized by long pulsation periods of greater than 100 days and high near-infrared and bolometric luminosities. In particular, they have large amplitude variations in infrared and visual wavelengths. Mira variables eject a considerable portion of their mass into surrounding regions, due to their pulsation, and the mass of the resulting circumstellar dust shells is correlated to their periods \citep{1993A&A...273..570A}. Therefore, while all stars experience extinction due the intervening interstellar dust, Miras also experience intrinsic extinction due to circumstellar dust, and this latter phenomenon affects longer period stars the greatest.
	
	AGB variables lie on distinct sequences in diagrams of period versus luminosity, with each sequence corresponding to a different normal mode of pulsation. Mira variables lie on a single sequence, which corresponds to the fundamental mode \citep{1996MNRAS.282..958W,2015MNRAS.448.3829W}. While these sequences are not as tight as those of some other types of variables, most notably Cepheids, they are still quite well-defined. For example, \citet{2015AJ....149..117M} measured a root-mean-square of 0.087 for the $K_{s}$-band near-infrared period-luminosity relation of Large Magellanic Cloud (LMC) fundamental-mode Cepheids, \citet{2017AJ....154..149Y} measured a scatter of 0.118 mag to the period-luminosity relation of the oxygen-rich Miras in the LMC with periods shorter than 400 days. 
	
	The Mira period-luminosity relations can be expressed in the form $M_{Ks} = \delta + \rho [\log P - 2.38]$, where the period $P$ is measured in days. For example, \cite{2008MNRAS.386..313W}, who used a sample of LMC Miras, measured the slope to be $\rho = -3.51 \pm 0.2$, while the zero-point, which was derived using solar-neighbourhood Miras, was determined to be $\delta = -7.15 \pm 0.07$, assuming an extinction-corrected LMC distance modulus of $\mu_{LMC}= 18.39 \pm 0.05$ \citep{2007MNRAS.379..723V}. Thus, the parameters of the Mira period-luminosity relation can be calculated to better than 6\% uncertainty. More recently, \citet{2017AJ....154..149Y} measured $M_{Ks} = \mu_{LMC} + (-7.23 \pm 0.001) + (-3.77 \pm 0.07) [\log P - 2.38]$, which is written in Table 3 of that work as $K_{s} =  (-6.93 \pm 0.001) + (-3.77 \pm 0.07) [\log P - 2.30]$.
	
	If the total extinction of the Miras' light is known, then the period-luminosity relation makes Mira variables useful distance indicators, since
	\begin{equation}
	\mu = m - M - A \label{eqn:distmod}
	\end{equation}
	where $\mu$ is the distance modulus, $m$ is the apparent magnitude, $M$ is the absolute magnitude, and $A$ is the amount of extinction. One goal of this paper is to determine the viability of this technique, since a number of difficulties can arise in making such a distance estimation. First, it is important to select a sample of Mira variables that has sufficient photometric completeness. Nearer stars tend to be brighter, while distant stars tend to be fainter; therefore, if the sample contains a disproportionate amount of bright or faint stars, there may be a bias in the distance determinations. Secondly, the local Galactic Mira period-luminosity relation determined by \cite{2008MNRAS.386..313W} has a slope and zero-point based on Miras from the LMC; however, different galaxies vary widely in age and metallicity. The dependence of the period-luminosity relation on properties such as age and chemical composition has not been probed in great depth, so measurements of distances to other galaxies that rely on an LMC-based period-luminosity relation may require a correction based on these differences. Therefore, another goal of this paper is to use theoretical models to determine whether such corrections are needed.
	
	We use a complete sample of Miras and improved extinction estimates to measure the distance to the Galactic centre. The Galactic centre provides a useful testbed for using Miras as distance indicators, as its mean distance is measured precisely, and the characteristics of the stellar population in the surrounding Bulge are well-measured. Currently, the best estimate of the distance is $R_0 = 8.122 \pm 0.031$ kpc \citep{2018A&A...615L..15G}, which comes from modelling the astrometric and radial velocity time-series data of the orbit of the star S2 around the supermassive black hole in the Galactic centre. This result is consistent with the prior literature value of $R_0 = 8.2 \pm 0.1$ kpc \citep{2016ARA&A..54..529B}, which was determined by examining distance measurements made using a variety of techniques; however, there is tension between these values and a recent measurement of 8.9 kpc made using Miras \citep{2016MNRAS.455.2216C}. In this paper, we calculate our own distance estimate, as well as examine issues contributing to the uncertainty in this measurement. 
	In Section \ref{sec:Data}, we select a photometrically complete sample of Bulge Miras and state the assumptions we make about the Bulge in fitting the distance to the Galactic centre. In Section \ref{sec:Extinction}, we compare the results of measuring distance using different extinction estimates. In Section \ref{sec:Models}, we examine the dependence of the Mira period-luminosity relation on age, helium abundance, and metallicity using theoretical models of Mira variables. We conclude our results in Section \ref{sec:Conclusion}.
	
	\section{Data and Bulge Model Assumptions} \label{sec:Data}
	
	\subsection{Distance measurement method and assumptions}\label{sec:distmethod}
	
	Our intent is to re-examine the distance measurement to the Galactic centre using Mira variables in the Galactic bulge. This is accomplished by plotting the distance modulus of each star in our sample against Galactic longitude, then applying a least-squares fit to the sample and choosing the zero-point of the solution as the Galactic centre \citep{2006ApJ...651..197C}. To accurately determine $R_0$, it is important that we choose a photometrically complete sample of Miras. We begin by using 643 Mira variables selected by \citet{2016MNRAS.455.2216C} from two fields observed by the South African Astronomical Observatory (SAAO), 6528 Miras listed by \cite{2013AcA....63...21S} from the Optical Gravitational Lensing Experiment (OGLE), and 1286 Miras from the MACHO survey \citep{2013OEJV..159....1B, 2016arXiv161200706B, 1999PASP..111.1539A}. We assume these stars are oxygen-rich, since nearly all Bulge Miras are O-rich and contamination by carbon-rich Miras would predominantly occur at very long periods, which we address in section \ref{sec:compare}. In addition, any error introduced by such contamination would not be caused directly by the presence of C-rich Miras, but rather by the difference in contamination rate by C-rich Miras between the Bulge and the LMC \citep{2008MNRAS.386..313W}.
	
\begin{figure*}
		\centering
		\includegraphics[width=0.65\textwidth]{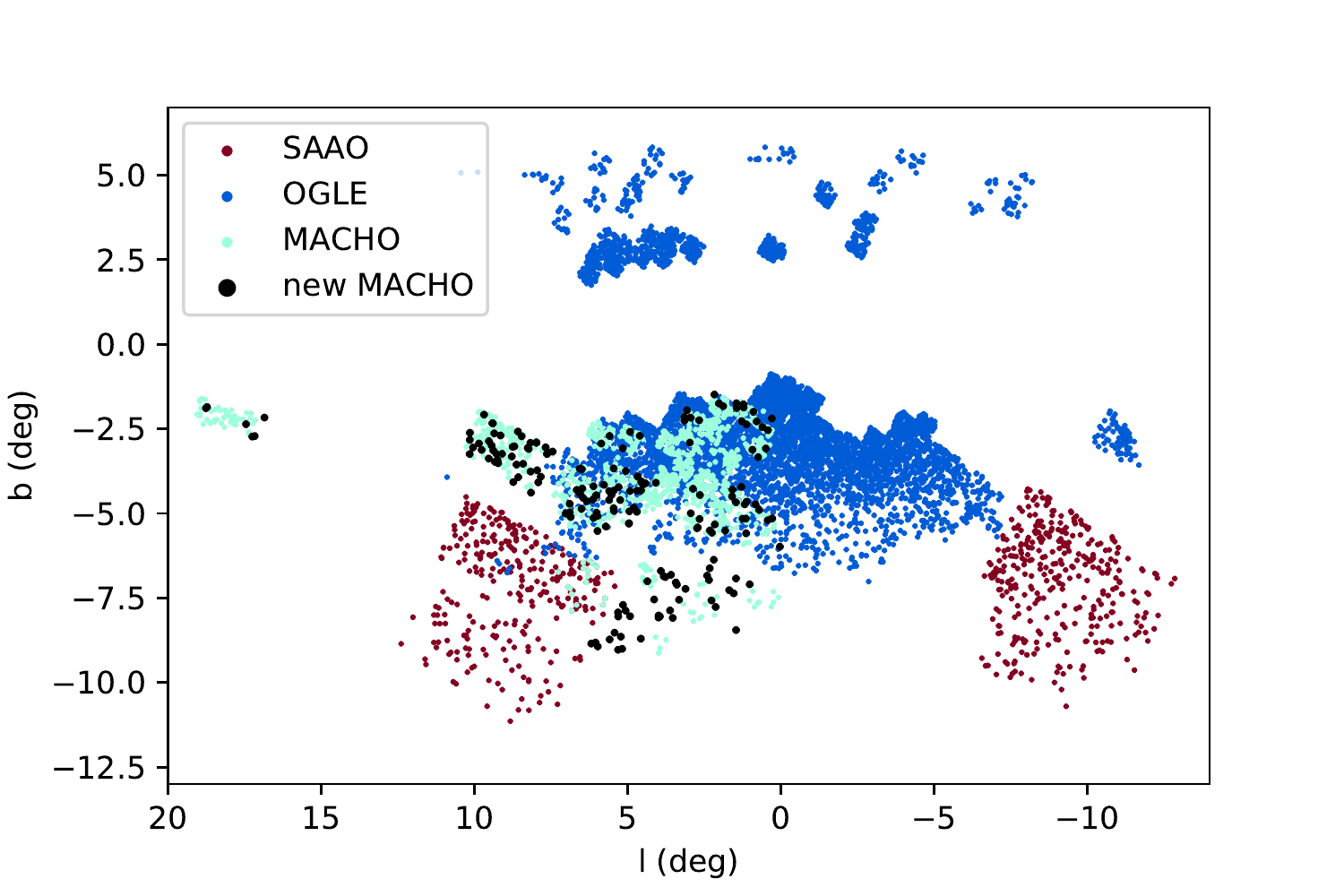}
		\caption{Map in Galactic coordinates of all the Miras used in the study. Each Mira is color-coded according to the survey that it came from. The newest 192 Miras identified in the MACHO survey are marked in black dots.}
		\label{fig:skymap}
	\end{figure*}

	We cross-match the coordinates of each star with the Wide-field Infrared Survey Explorer, or WISE \citep{2010AJ....140.1868W}, using VizieR with a match radius of 0.5 arcseconds in order to obtain the $W1$ and $W2$ magnitudes. This leaves 635 SAAO Miras, 5821 OGLE Miras, and 1238 MACHO Miras. The distribution of the Miras in Galactic coordinates is shown in Figure \ref{fig:skymap}. We also cross-match all the Miras with the Two Micron All-Sky Survey (2MASS) to obtain the $JHK_s$ photometry for our sample \citep{2006AJ....131.1163S}. The 2MASS $JHK_s$ arrays observed objects nearly simultaneously, so that all photometric measurements are made at the same phase in the Miras' luminosity variation. Both the WISE and 2MASS magnitudes are calibrated to the Vega scale.
	
	Some of the equations imported for our analysis were derived using SAAO magnitudes. There are no equivalents to these equations that are based on 2MASS measurements, so to make these equations compatible with the 2MASS photometry, we invert the transformations given by \cite{2001AJ....121.2851C} and revised at `\hyperlink{www.astro.caltech.edu/~jmc/2mass/v3/transformations/}{www.astro.caltech.edu/$\sim$jmc/2mass/v3/transformations/}':
	\begin{equation}
	\begin{aligned}
	(K_s)_{2MASS} &= K_{SAAO} - 0.024 + 0.017 (J-K)_{SAAO} \\
	(J-H)_{2MASS} &= 0.944 (J-H)_{SAAO} - 0.048 \\
	(J-K_s)_{2MASS} &= 0.944 (J-K)_{SAAO} - 0.005 \\
	(H-K_s)_{2MASS} &= 0.945 (H-K)_{SAAO} - 0.043.
	\end{aligned}
	\label{eqn:transform}
	\end{equation}
	 All values and relations given from this point forward are in or have been converted to the 2MASS system, unless otherwise specified. Unfortunately, those relations are largely calibrated off stars bluer than Miras. The sample includes 635 stars for which there are both SAAO and 2MASS measurements. We find that the mean offset between the SAAO-derived versus measured 2MASS magnitudes are 0.01, 0.00, and 0.03 mag in $J,H,$ and $K_{s}$. We thus opt not to adjust the color transformations. 
	
	We identify and remove duplicate stars between catalogues if the difference in position is less than $0.001 \degree$ and the periods differ by less than $50$ days. We find 183 such duplicate stars, which leaves a total of 7511 Miras for analysis. For a sense of the size of the distribution of Miras across the sky, the sample extends a mean angle of $5.3\degree$ from the Galactic centre, corresponding to an average transverse separation of about 0.7 kpc. If we only include SAAO and MACHO objects in this calculation, then the mean angle increases to $8.2\degree$, corresponding to an average transverse separation from the Galactic centre of 1.1 kpc.
	
	For comparing the color predictions of computational models used in this study to real stars, we also obtain 251 solar-neighbourhood Miras and semi-regular variables from Table 1 of \cite{2000MNRAS.319..728W}, as these stars do not suffer from significant interstellar extinction (see Section \ref{sec:Models}). These stars were originally observed by \textit{Hipparcos} \citep{1997ESASP1200.....E}, so we obtained the period for each object from the \textit{Hipparcos} data. There are 63 matching objects in the WISE catalogue, which gives $W1$, $W2$, and $W3$ magnitudes. There are also 52 matching objects in the SDSS catalogue \citep{2012ApJS..203...21A}, which gives $ugriz$ magnitudes on the AB scale.
	
	\subsection{Photometry and period data quality}\label{sec:photometry}
	
	To assess the photometric completeness of each survey, we make the following assumptions. If we wish to sample stars on both the near and far sides of the Bulge, then we require our data to cover distances between 4 and 12 kpc. In addition, \cite{2008MNRAS.386..313W} find that Miras with periods of $2.1 < \log P < 2.7$ have a range of absolute $K_s$-magnitudes between -6.3 and -8.4. Estimating an average extinction of 0.75 and plugging these values in equation \ref{eqn:distmod}, we find that to have a Mira sample that is complete on both sides of the bulge, our sample must have apparent $K_s$-magnitudes at least as bright as 4.6 and at least as faint as 9.8. Since the range in $K_s$ for any Mira is greater than 0.4 magnitudes and our data is comprised of single observations, our analysis would be improved with more time-series observations of the Bulge. At the same time, the light curves of Miras are fairly symmetric in the infrared, and so should not be biased \citep{1995MNRAS.273..383G}.
	
\begin{figure*}
		\centering
		\subfigure{\label{fig:Kmag_SO}\includegraphics[width=55mm]{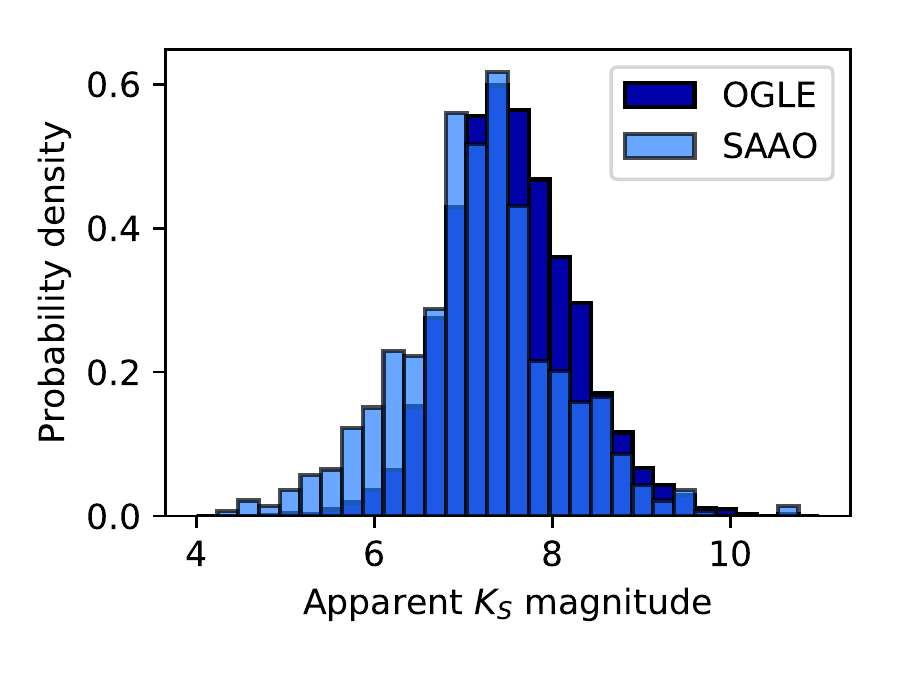}}
		\subfigure{\label{fig:Kmag_OM}\includegraphics[width=55mm]{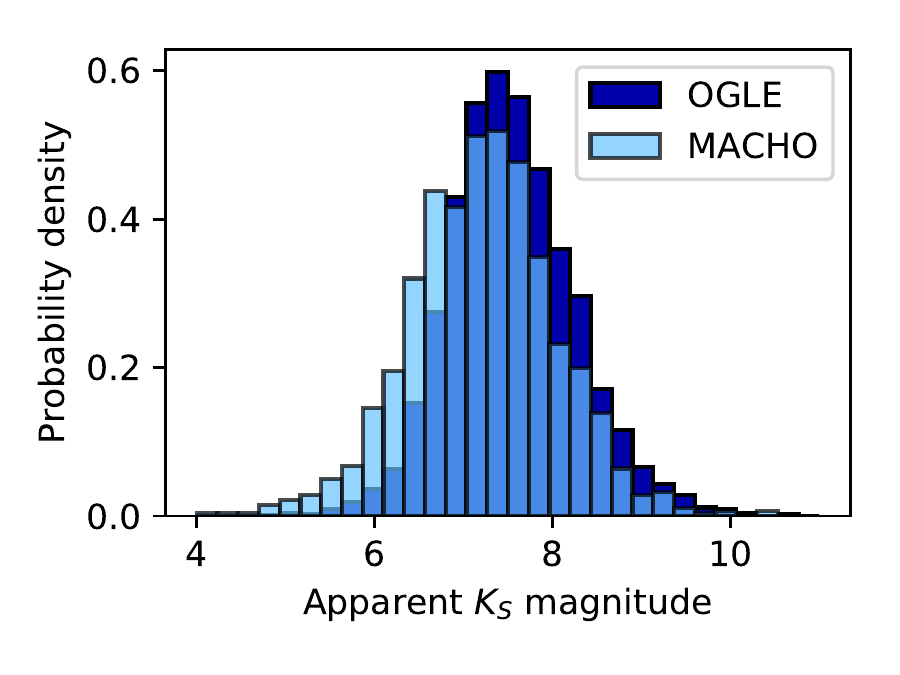}}
		\subfigure{\label{fig:Kmag_SM}\includegraphics[width=55mm]{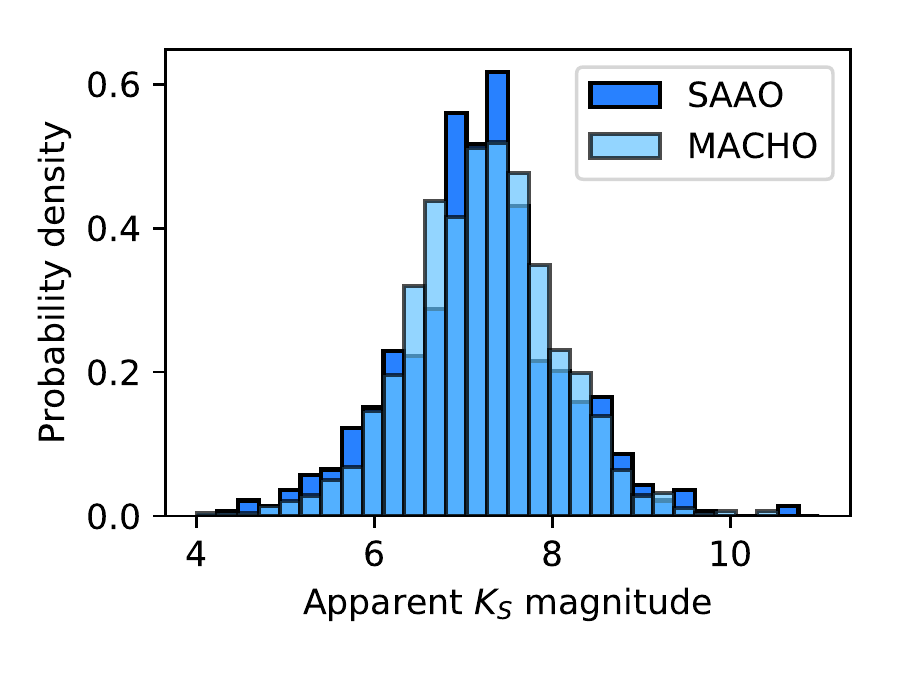}}
		\caption{$K_s$ magnitude distributions of Miras in various catalogues. Each histogram is normalized such that the height represents a probability density. In the left panel, there is an offset between the catalogues that indicates the OGLE survey is biased towards fainter stars compared to the SAAO survey. In the center panel, the MACHO survey also has a larger left tail, indicating it contains brighter stars than the OGLE survey. In comparison, the right panel shows that the distributions for SAAO and MACHO almost completely overlap.}
		\label{fig:Kmag}
	\end{figure*}

	We compare distributions of the apparent 2MASS $K_s$ magnitudes for the Bulge Miras in each catalogue in Figure \ref{fig:Kmag}. While the SAAO and MACHO Miras show nearly identical $K_s$ magnitude distributions, the OGLE catalogue stars appear shifted towards higher magnitudes, indicating that the survey is comparatively faint. This is not due to incompleteness at the faint end in the SAAO and MACHO samples. As stated previously, to be considered photometrically complete, each sample must cover at least the apparent $K_s$-magnitudes $4.6 < m_K < 9.8$. Each catalogue is sampled at the faint end and contains stars with $m_K > 10$; however, while both SAAO and MACHO have stars brighter than $m_K = 4.5$, the brightest Mira in the OGLE catalogue is only $m_K = 4.9$. In addition, it has been shown that the MACHO survey is relatively complete in Bulge RR Lyrae stars, which have an average absolute magnitude in the $V$ band of $\overline{M_V} \sim 0$ \citep{2008AJ....136.2441K}. According to Table 4 of \cite{2002A&A...384..925K}, Miras are typically brighter, with an average absolute magnitude between $-3.5 < \overline{M_V} < -1$, depending on their periods. This implies that the Miras from the MACHO and SAAO catalogues are well-sampled throughout the Bulge.
	
	If the SAAO and MACHO surveys are as complete as OGLE on the faint end, then the relative shift of the OGLE survey towards higher magnitudes means that the OGLE survey does not contain many of the brighter stars on the near side of the Bulge. This is due to the saturation limit of the OGLE survey at $I \approx 12.5$ mag \citep{2013AcA....63...21S}. Since only a small fraction of overexposed long-period variables are included in the catalogue, the OGLE Miras have low completeness for brighter stars. In addition, since the mean of the $I$-magnitude distribution for the OGLE Miras is $\bar{I} = 13.77 \pm 1.27$ magnitudes, there is some overlap with the saturation limit. We conclude that a photometrically complete sample of Miras, i.e. a set of Miras that is well sampled throughout the Galactic Bulge, should not contain OGLE stars, which leaves us with 1863 Miras in our sample. Further justification of this is presented in Section \ref{sec:compare}.
	
	In examining the quality flags associated with the 2MASS photometry for the Bulge Miras, we find that less than 3\% of the 2MASS measurements are flagged as unreliable or of poor quality. In addition, we find that 77.1\% of the WISE measurements for the Bulge are unaffected by artefacts, while 18.5\% may be contaminated by scattered light and 4.0\% contaminated by diffraction spikes from nearby bright sources, with the remaining 0.4\% contaminated by other artefacts. Since the stars affected by 2MASS artefacts constitute only a small fraction of our data and the WISE photometry does not play a central part of the analysis, we have chosen to use the full set of measurements, as the contaminated measurements should not significantly impact our results.
	
	The reliability of the periods determined by SAAO, OGLE, and MACHO is demonstrated by comparing the periods of the duplicate stars discussed previously. If we examine the ratios of the duplicate star periods, we get a mean period ratio of $P_1/P_2 = 1.00 \pm 0.02$, which verifies the reliability of our period data. The \textit{Hipparcos} periods of the solar-neighbourhood Miras that we use have previously been shown to be consistent with other determinations \citep{2000MNRAS.319..728W}.

	\subsection{Bulge model}\label{sec:assumptions}
	The distance modulus for each star is given by equation (\ref{eqn:distmod}). The absolute magnitude in the $K_s$-band is given by the period-luminosity relation in \cite{2008MNRAS.386..313W}, which we have converted to the 2MASS system:
	\begin{equation}
	M_{Ks} = -3.50(\log P - 2.38) - 7.257. \label{eqn:whitelock_pl}
	\end{equation}
	In this equation, $M_{Ks}$ is the mean magnitude. This relation was transformed by using the first of equations (\ref{eqn:transform}) and substituting the $(J-K)_{SAAO}$ color in that equation with $J-K = -0.39 + 0.71\log P$, which was derived by \cite{2000MNRAS.319..728W}.	The zero-point is 0.1 magnitudes brighter than the best-fitting value that \cite{2008MNRAS.386..313W} derived from LMC data, as we are using an updated distance modulus to the LMC of 18.49, which is precisely and accurately measured from eight long-period, late-type eclipsing systems composed of cool, giant stars \citep{2013Natur.495...76P}. Interestingly, this value is in agreement with that derived by \cite{2008MNRAS.386..313W} using the \textit{Hipparcos} parallaxes of solar neighbourhood Miras to anchor their distance scale. The uncertainties in equation (\ref{eqn:whitelock_pl}) are discussed in Section \ref{sec:Zeropoints}. We note that the distances derived in this manner are nearly indistinguishable from the distances calculated using the \cite{2017AJ....154..149Y} period-luminosity relation (the offsets are typically no more than 30 pc).

	\begin{figure}
		\centering
		\includegraphics[width=0.42\textwidth]{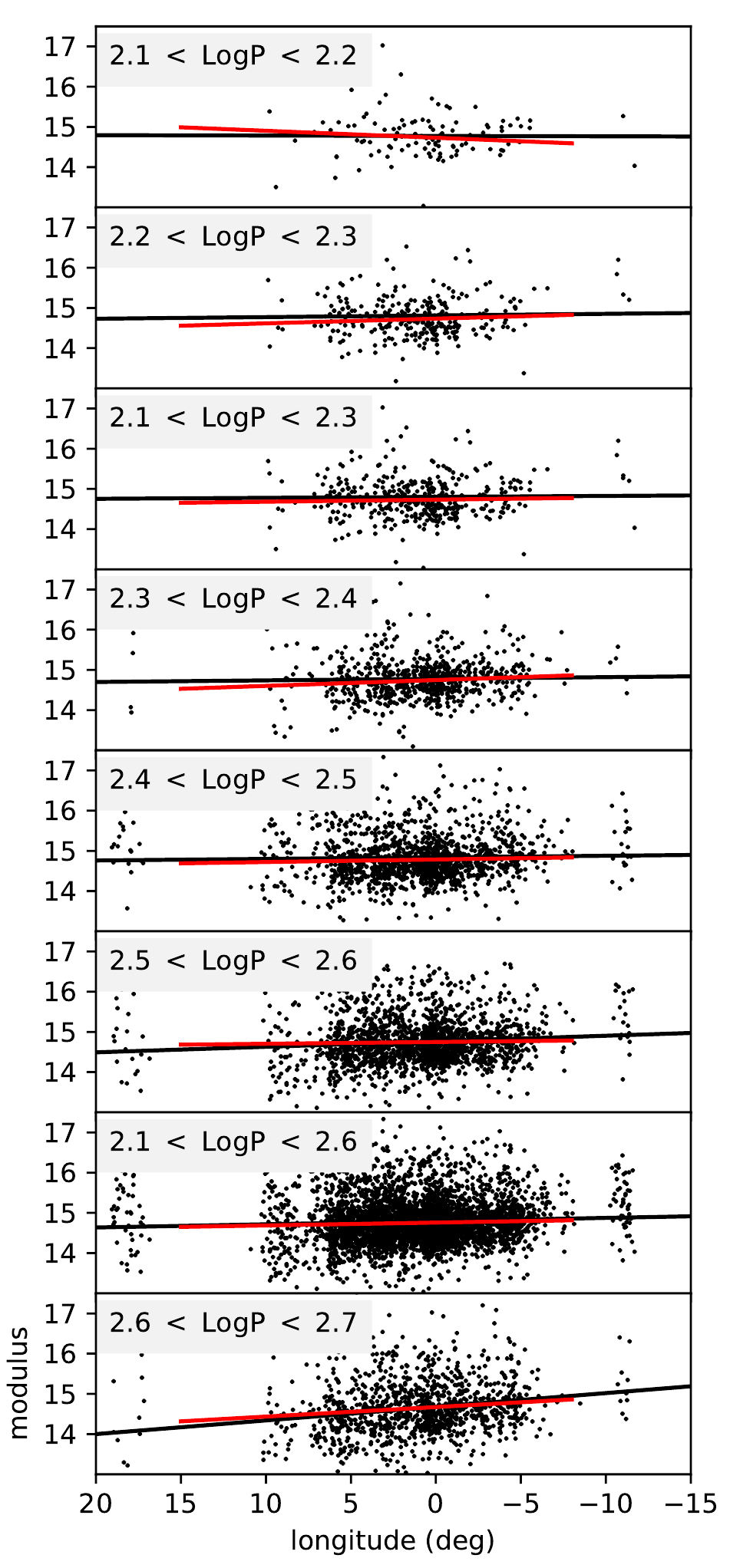}
		\caption{Distance modulus versus longitude for all stars (including OGLE) of different period intervals and $|b| < 4.5\degree$. The lines represent least-squares fits, with the black line being the fit through all the stars and the red line being the fit through stars with $|l| < 3\degree$. The zero-points of the fits are given in Table \ref{tab:dist_reprod}.}
		\label{fig:linefit}
	\end{figure}
	
	For each method of estimating extinction, the distance to the centre of the Galaxy, $R_0$, is estimated by computing the least-squares fit of distance modulus versus longitude, and then taking the distance at $l = 0^{\circ}$ as the best fit \citep{2006ApJ...651..197C}, as shown in Figure \ref{fig:linefit}. Although distance modulus is a logarithmic quantity, the variation in $\mu$ is much smaller than the mean value of $\mu$, which allows us to approximate the distance modulus as a linear quantity. The inclination of the modulus-longitude relation at longer-periods is due to the bar structure of the Bulge. The nearer side of the bar is at positive longitudes, and thus stars at negative longitudes appear fainter on average \citep{2006ApJ...651..197C,2016MNRAS.455.2216C}. 
	
	The fact that our sample of stars does not cover the Galactic midplane allows us to avoid selection effects caused by the highly filamentary structure of extinction near $|b| = 0$ \citep{2016ApJ...818..130B}. However, the sample's asymmetric distribution in $b$ does affect our derivation of the distance modulus, since this introduces a dependence on Galactic latitude \citep{2015MNRAS.450.4050W}. This effect is due to the elliptic shape and angle of the Bulge. In fact, \cite{2016MNRAS.456.2692N} fit the models of \cite{2015MNRAS.450.4050W} to get the following equation for the apparent magnitude of red clump stars in the Bulge
	\begin{equation}
	I_{RC,0} = 14.3955 - (0.0239 \times l) + (0.0122 \times |b|),
	\end{equation}
	which indicates that the expected dependence of distance modulus on latitude is 0.0122 magnitude per degree. The mean absolute latitude of the SAAO and MACHO stars is $4.84 \degree$, so this gives an expected offset of 0.06 magnitudes. Assuming $\mu_0 \approx 14.6$, this leads to a distance offset of about 0.23 kpc, which is on the order of other sources of uncertainty. This indicates that the distance moduli we derive require a small geometric correction. We make this correction by subtracting $0.2$ kpc from the distance estimate. Alternatively, one might directly apply the distance modulus offsets from \citet{2015MNRAS.450.4050W} rather than the smoothed relation above. The final result differs by $\sim$0.007 kpc, and thus the correction is negligible. 
	
	Lastly, in order to remove foreground and background Miras (i.e., Miras that are not actually part of the Bulge population), we only use Miras that are within 4 kpc of the Galactic centre. To accomplish this, we assume the \cite{2016ARA&A..54..529B} distance to the Galactic centre of $8.2$ kpc and apply this spherical volume restriction separately for each extinction estimate, since each estimate gives different distances to the stars. In each case, between $12-17\%$ of objects were removed from the dataset to be fit ($6\%$ in the case of the \citealt{2011A&A...534A...3G,2012A&A...543A..13G} map, where stars outside of a certain range of angles were not assigned reddening values, see Section \ref{sec:dust}). The stars that lay outside of the sphere are closer to the plane on average than stars inside the sphere, which is consistent with the removed stars being part of the background disc.

	\section{Extinction Corrections}\label{sec:Extinction}
	
	When studying Miras, there are two types of extinction one must take into account. The first is extinction caused by interstellar dust. This is an effect experienced by all types of stars, and the amount of extinction depends on the direction of and distance along the line of sight. The second type of extinction is caused by dust expelled from the Miras, and thus only occurs in the regions near the star. Most extinction maps only measure interstellar extinction; however, there are methods for measuring the total extinction from both interstellar and circumstellar dust and we are careful to make this distinction \citep{2017AJ....154..149Y}.
	
	In our study, we compare several different methods of estimating extinction. We use the intrinsic period-color relations given by \cite{2000MNRAS.319..728W} and \cite{1995MNRAS.273..383G} to predict total extinction. We use the reddening map described by \cite{1998ApJ...500..525S} and recalibrated by \cite{2011ApJ...737..103S}, and the reddening map given by \cite{2011A&A...534A...3G,2012A&A...543A..13G} to measure interstellar extinction only. We also use the Rayleigh-Jeans color excess method \citep{2011ApJ...739...25M}, which relates extinction to mid-infrared colors.
	
	The reddening maps of \cite{2012A&A...543A..13G} and \cite{2011ApJ...737..103S} are "2D" reddening maps, meaning they do not account for the increase in extinction along the line of sight, but rather are mean reddening values for small angular regions on the sky. The error in the extinction estimate from this depth effect is expected to be small, as a typical source from our sample is $\sim$700 pc removed from the Galactic midplane, with none closer than $\sim$200 pc. Since the scale height of dust in the Milky Way is about 125 pc, our sources are far enough from the midplane that the prevalence of dust and, therefore, the dependence of extinction on distance is reduced \citep{2006A&A...453..635M}.

	\subsection{Period-color Relations} \label{sec:WG}
	
	\cite{2000MNRAS.319..728W} derive the following relations between the periods and mean intrinsic colors for local Galactic Miras, which has been transformed to give colors on the 2MASS system:
	\begin{equation}
	(J-K_s)_0 = -0.37 + 0.67 \log P. \label{eqn:whitelock_map_saao}
	\end{equation}
	Similarly, \cite{1995MNRAS.273..383G} give the relation
	\begin{equation}
	(J-K_s)_0 = -0.12 + 0.53 \log P \label{eqn:glass_map_saao}.
	\end{equation}
	These equations are plotted in Figure \ref{fig:PC_schl+wg}. By subtracting these relations from the observed Mira colors, we can obtain the reddening of the Miras' light, or the color excess in $(J-K_s)$. Reddening and extinction are correlated, since both are caused by dust; however, it is easier to first predict reddening instead of extinction, since an object's intrinsic color is not affected by its distance. The extinction in $K_s$ can then be derived from the color excess in $(J-K_s)$ using either the total-to-selective extinction ratio determined by \cite{2009ApJ...696.1407N} from Bulge red clump and red giant branch stars:
	\begin{equation}
	\frac{A_{K_s}}{E(J-K_s)} = 0.495,
	\end{equation}
	or that given by \cite{1999hia..book.....G}:
	\begin{equation}
	\frac{A_{K_s}}{E(J-K_s)} = 0.53.
	\end{equation}
	Combining these equations and extinction coefficients gives four methods of estimating extinction. Since the period-color relations predict the intrinsic colors of the Miras, they automatically account for both interstellar and circumstellar reddening, making them total reddening maps. The error due to circumstellar dust is likely reduced -- whereas Galactic dust maps effectively set the total-to-selective extinction ratio of circumstellar dust to zero, estimates from period-color relations set them equal to the total-to-selective extinction ratios of interstellar dust. It would be advantageous if robust estimates of circumstellar extinction coefficients were available. This would improve distance estimates, and provide the option of extending the analysis to longer-period Miras, which have more circumstellar dust. 
	
	We find that the \cite{2000MNRAS.319..728W} relation gives negative reddening values for 15 objects and the \cite{1995MNRAS.273..383G} relation gives negative values for 3 objects. Whether we choose to keep these objects, remove them, or replace them with reddening values from another map such as \cite{2011ApJ...737..103S} (see Section \ref{sec:dust}), we find that the $R_0$ distances we derive differ by less than 0.002 kpc for stars of period $\log P < 2.6$ and 0.03 kpc for stars of period $\log P > 2.6$. As our final distance estimate will be restricted to stars with $\log P < 2.6$ (justified in Section \ref{sec:dust}), this correction is negligible.  
	
	\subsection{Dust Reddening Maps} \label{sec:dust}
	
	The map described by \cite{1998ApJ...500..525S} was recalibrated by \cite{2011ApJ...737..103S} using stellar spectra from the Sloan Digital Sky Survey \citep{2011ApJS..193...29A}. The \cite{2011ApJ...737..103S} map predicts reddening caused by interstellar dust, which is therefore an underestimate of the reddening to our sample, which is often contaminated by circumstellar dust. For this map, we use the larger \cite{1999hia..book.....G} coefficient to convert the reddening to extinction. 
	
\begin{figure}
	    \hspace{-5mm}
		\includegraphics[width=0.55\textwidth]{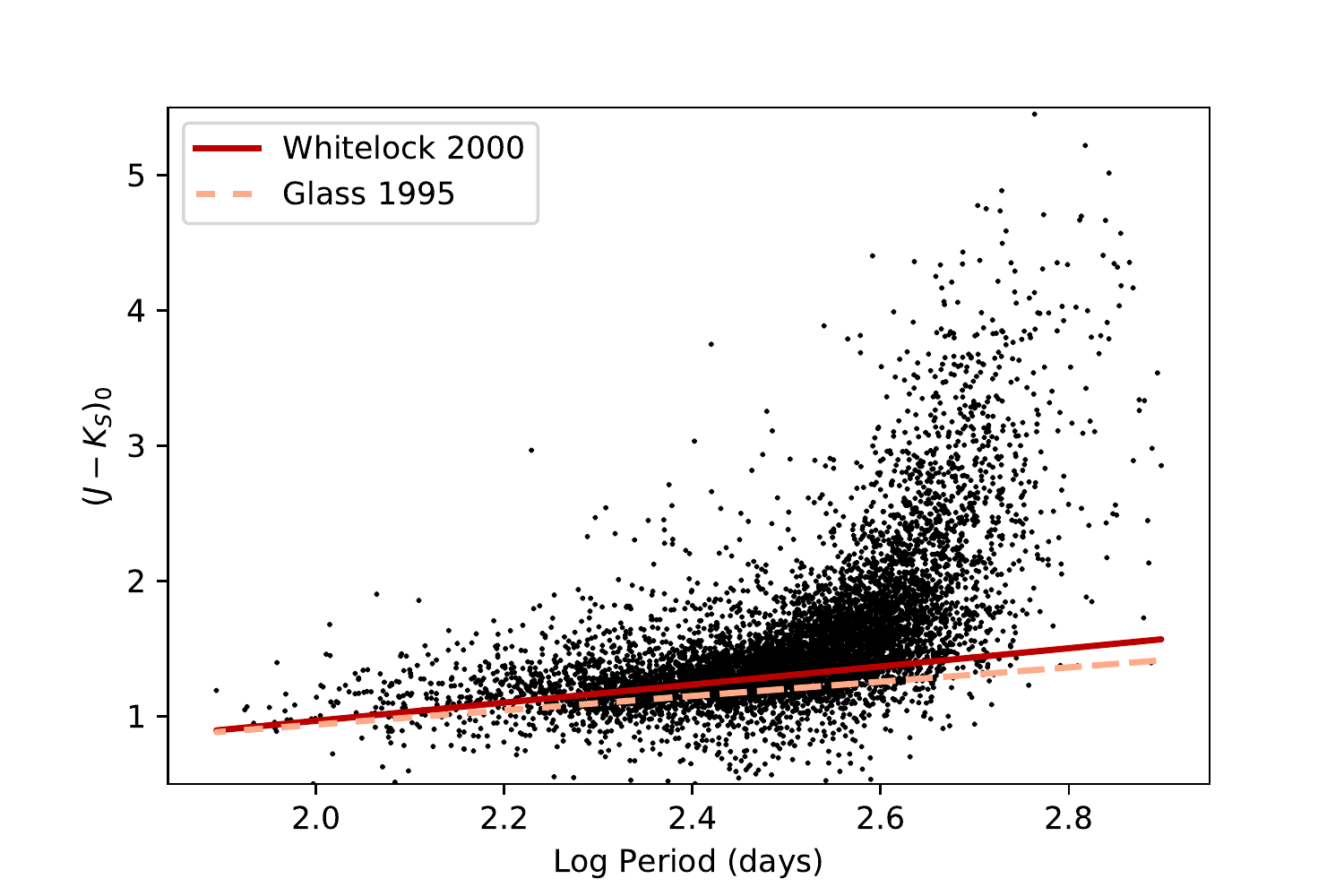}
		\caption{A color-period diagram of all the Miras, corrected for extinction using the \protect\cite{2011ApJ...737..103S} reddening map. The lines represent the \protect\cite{2000MNRAS.319..728W} and \protect\cite{1995MNRAS.273..383G} period-color relations. Because the \protect\cite{2011ApJ...737..103S} map only takes into account the interstellar extinction, the data points deviate from the reported period-color relation due to circumstellar extinction. Nonetheless, at short periods ($\log P < 2.6$) the lines track the data well. At $\log P > 2.6$, circumstellar extinction dominates.}
		\label{fig:PC_schl+wg}
	\end{figure}

	Figure \ref{fig:PC_schl+wg} shows the $(J-K_s)$ colors of all Miras dereddened using the \cite{2011ApJ...737..103S} map, with lines drawn in to represent the period-color relations given by \cite{2000MNRAS.319..728W} and \cite{1995MNRAS.273..383G}. Up to about $\log P \approx 2.6$, the data follows these lines quite well; thus, for Miras of shorter periods, the \cite{2011ApJ...737..103S} map appears to be in good agreement with the \cite{2000MNRAS.319..728W} relation converted to 2MASS photometry. Longer period Mira variables have higher mass loss rates, resulting in greater amounts of circumstellar dust, so the discrepancies between the maps at values of $\log P \gtrsim 2.6$ are expected \citep{1990ASPC...11..365W,1993A&A...273..570A}. For stars with $\log P < 2.6$, the \cite{2000MNRAS.319..728W} reddening estimate is larger than the \cite{2011ApJ...737..103S} estimate by a median offset of 0.051 magnitudes, indicating the two extinctions maps are consistent with each other up to $\log P \approx 2.6$.
	
\begin{figure}
    \centering     
    \subfigure{\label{fig:excess_whitgon}\includegraphics[width=70mm]{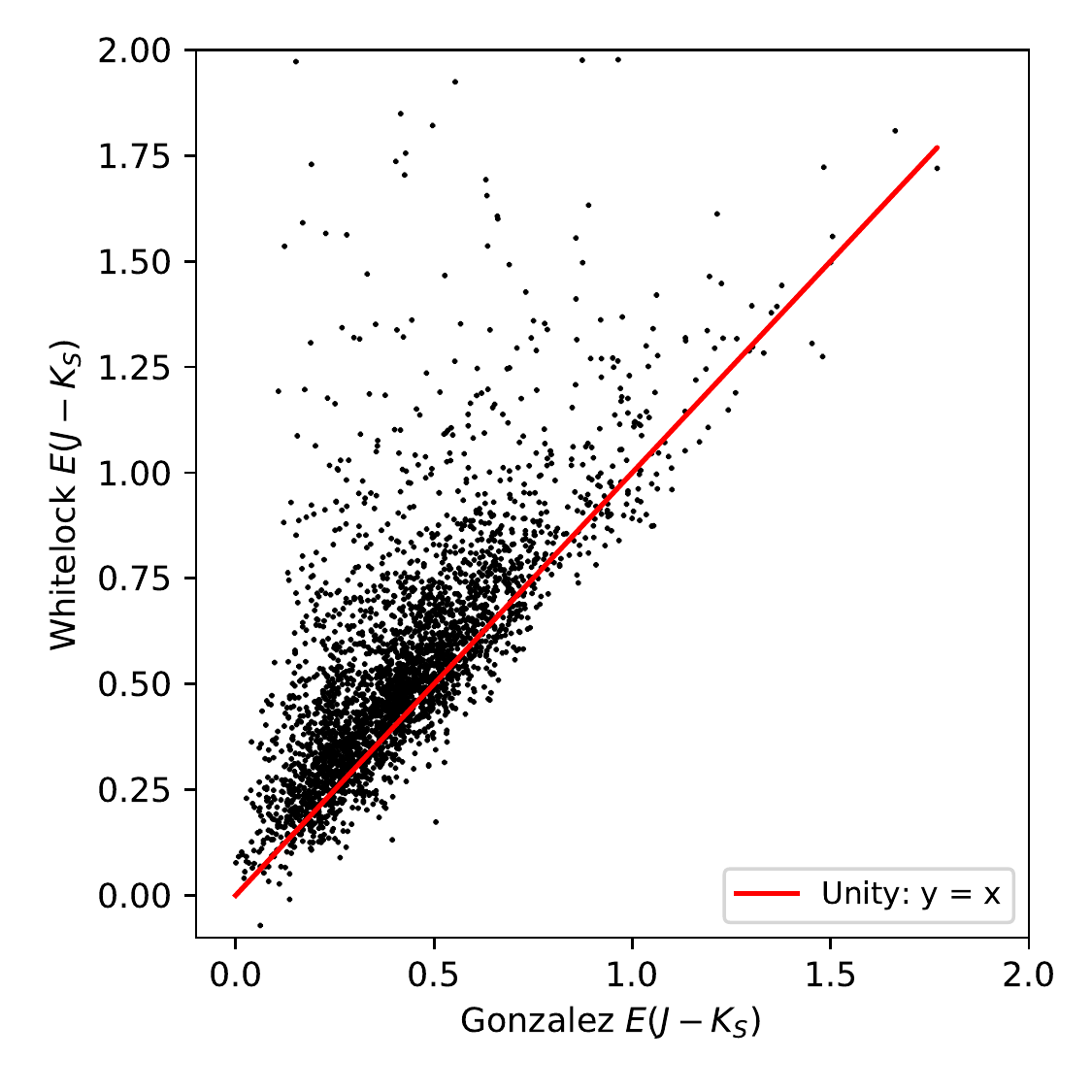}}
    \subfigure{\label{fig:excess_schlgon}\includegraphics[width=70mm]{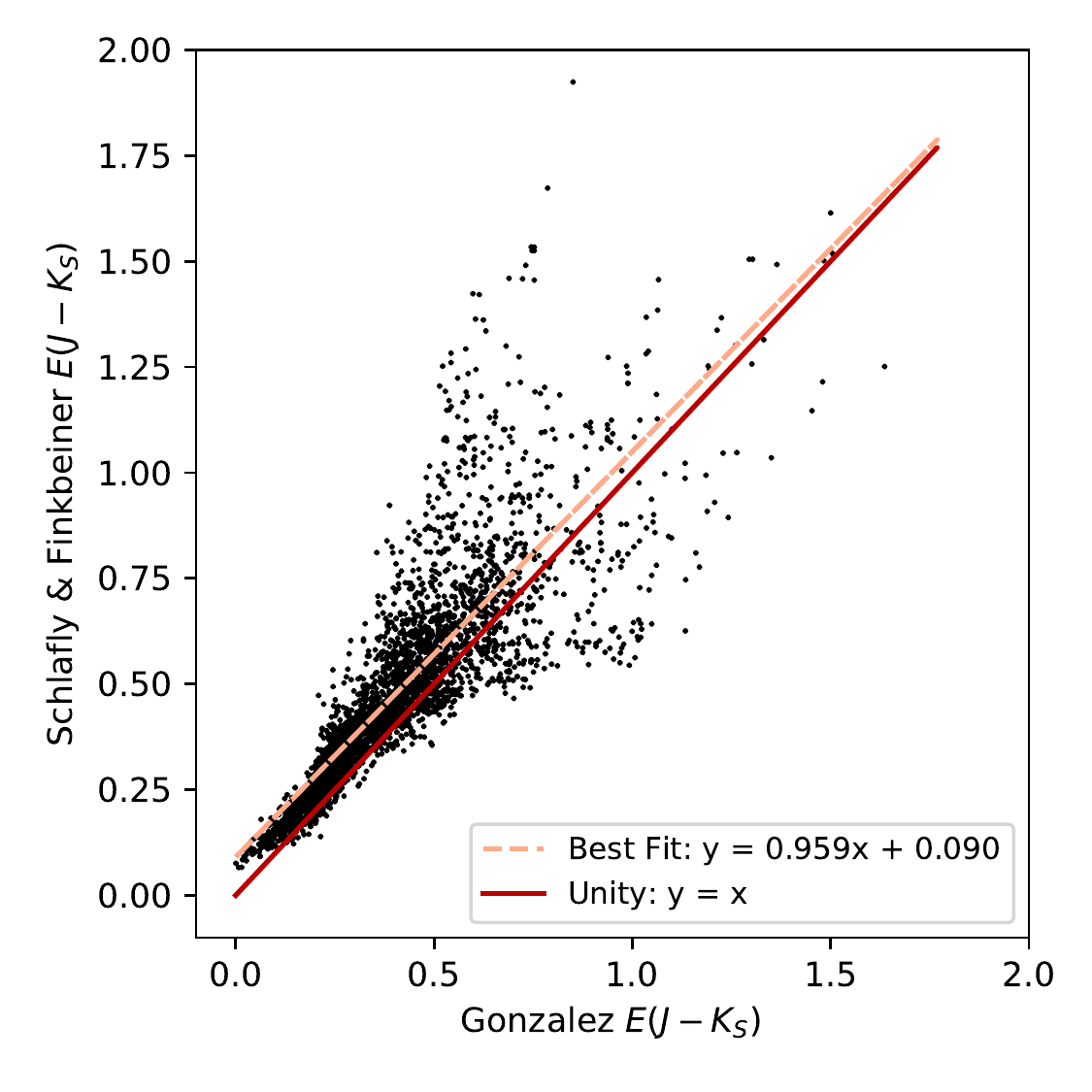}}
    \caption{The top panel shows extinction measured according to the method using the \protect\cite{2000MNRAS.319..728W} relation versus extinction according to the \protect\cite{2012A&A...543A..13G} map, for stars with $\log P < 2.5$. The data points above the line $y=x$ are caused by the fact that the \protect\cite{2000MNRAS.319..728W} method is a total extinction map, while the \protect\cite{2012A&A...543A..13G} map only predicts interstellar extinction. The bottom panel shows extinction measured according to the \protect\cite{2011ApJ...737..103S} map versus extinction according to the \protect\cite{2012A&A...543A..13G} map, for stars with $\log P < 2.5$. The data approximately follows the line of slope unity, indicating that the two extinction estimates are consistent with one another.}
\end{figure}

	The reddening map described by \cite{2011A&A...534A...3G,2012A&A...543A..13G} was determined using data from the ESO public survey, \textit{Vista Variables in the Via Lactea}. Similar to the \cite{2011ApJ...737..103S} extinction map, this map predicts interstellar extinction only, so we paired it with the \cite{1999hia..book.....G} ratio. The \cite{2011A&A...534A...3G,2012A&A...543A..13G} map covers the region $-10\degree \leq l \leq +10.2\degree$ and $-10\degree \leq b \leq +5\degree$, so we could not obtain reddening estimates for several of the stars in our sample. Figure \ref{fig:excess_whitgon} compares the \cite{2000MNRAS.319..728W} relation and \cite{2012A&A...543A..13G} calculation for the color excess in $(J-K_s)$. While the data shows the two reddening estimates are generally consistent with one another, there is significant scattering above the line of slope unity. This is, again, due to the fact that the method using the \cite{2000MNRAS.319..728W} relation estimates both interstellar and circumstellar extinction, while the \cite{2012A&A...543A..13G} map only accounts for interstellar extinction. 
	Figure \ref{fig:excess_schlgon} compares the \cite{2011ApJ...737..103S} and \cite{2012A&A...543A..13G} extinction estimates, and the plot also shows that the values are generally consistent, since they follow a trend line with a slope of one.
	
\subsection{Rayleigh-Jeans color Excess Method} \label{sec:RJCE}
	
	The Rayleigh-Jeans color excess method described by \cite{2011ApJ...739...25M} gives the extinction in terms of $H$ and $4.5\,\mu m$ (or W2) magnitudes and assumes that the intrinsic $(H - [4.5\mu])_0$ color of the stars being used is 0.08 magnitude:
	\begin{equation}
	A_{Ks} = 0.918(H - [4.5\mu] - 0.08).
	\label{eqn:RJCE}
	\end{equation}
	
	However, this method yields a median star distance of 5.22 kpc. The cause of this underestimate is most likely that these stars have different mid-infrared colors than what is assumed by equation (\ref{eqn:RJCE}), showing that the Rayleigh-Jeans method of reddening estimation fails for Mira variable stars. That is as expected from the analysis and discussion of \citet{2011ApJ...739...25M}, as the $(H - [4.5\mu])_0$ color of M-stars is expected to be redder.  We therefore did not include this method for further analysis. This issue is revisited in Section \ref{sec:Models}.
	
	\subsection{Comparison of \texorpdfstring{$R_0$}{R0} Estimates} \label{sec:compare}
	
	We divide the entire sample of stars into subsets of various period intervals, because of the effect described previously of longer-period stars having greater circumstellar extinction. We then use the method described in Section \ref{sec:Data} to calculate $R_0$ for each dataset. As an example, Table \ref{tab:dist_reprod} lists various estimates of $R_0$ using stars in certain intervals of period and Galactic coordinates, with the extinction correction coming from the \cite{2000MNRAS.319..728W} method, which accounts for both interstellar and circumstellar extinction. Since this initial test of the linear fits includes OGLE objects, uses SAAO photometry, and does not restrict the sample to stars within a spherical volume, the values are in fairly close agreement with the $R_0$ estimates calculated by \cite{2016MNRAS.455.2216C}---there are small offsets due to our different cross-matching criteria with the 2MASS catalogue. 
	
	To verify that we should not include Mira variables from the OGLE catalogue, we perform the line fittings for each catalogue separately, using the \cite{2000MNRAS.319..728W} extinction map with the \cite{1999hia..book.....G} coefficient. The results are shown in Table \ref{tab:dist_compare}. The OGLE catalogue yields larger $R_0$ values for both short-period and long-period Miras, which is expected given that many of the Miras are brighter than OGLE's saturation limit. There is a small offset between the SAAO and MACHO stars, the cause of which is unclear.
	
	The results of performing the line fittings for each extinction map excluding the OGLE Mira variables are shown in Table \ref{tab:dist_ext_nogle}. In comparison with Table \ref{tab:dist_ext}, which shows the same line fittings but including OGLE Miras, we see that every method of extinction estimation yields a smaller distance by about 0.5 kpc. This confirms that using the OGLE photometry biases our distance estimate to much higher values. In addition, in Table \ref{tab:dist_ext_nogle} the distances determined by the period-color relations are smaller than those determined by the interstellar extinction maps, in part because the period-color relations apply to individual Mira variables, making them sensitive to even small amounts of circumstellar extinction. The period-color relations also have the added advantage of not being susceptible to the resolution and depth effects that are inherent in the \cite{2011ApJ...737..103S} and \cite{2012A&A...543A..13G} reddening maps. 
	
	Of the two period-color relations used, we believe that of \cite{2000MNRAS.319..728W} produces the most reliable distance estimate. It can be argued that for the purpose of this study, the period-color relation derived by \cite{1995MNRAS.273..383G} is superior to the \cite{2000MNRAS.319..728W} relation, since the former was determined using Bulge Miras and the latter using solar-neighbourhood Miras. However, the reddening according to the \cite{2000MNRAS.319..728W} map is more consistent with the \cite{2011ApJ...737..103S} and \cite{2012A&A...543A..13G} reddening maps at shorter periods (Figure \ref{fig:PC_schl+wg}). Therefore, pairing the \cite{2000MNRAS.319..728W} relation with the \cite{2009ApJ...696.1407N} coefficient, which is supported by \cite{2011ApJ...737...73F}, gives the most reliable extinction estimate. 
	
	The conversion from reddening to extinction deserves scrutiny, because there is no conclusively established ``best" reddening law for the bulge. \cite{2016MNRAS.456.2692N} demonstrated that the reddening law varies significantly from sightline to sightline. For example, their Figure 4 shows a spread of about 40\% in the measured values of $A_{I}/E(V-I)$. The distribution of reddening laws they show includes both the \cite{2009ApJ...696.1407N} and \cite{1999hia..book.....G} reddening coefficients. There is currently no map of reddening law variations throughout the bulge and we do not have enough colors to calibrate the extinction to reddening ratio for the Miras on a star-by-star basis. 
	
	In estimating the distance, we choose to use stars with $\log P < 2.6$, since distances estimated with longer period stars are affected by greater circumstellar extinction, as well as possible contamination by C-rich Miras, as described by \cite{2011MNRAS.412.2345I}, \cite{2017MNRAS.469.4949M}, and \cite{2017AJ....154..149Y}. The third dredge-up, which can produce C-rich Miras, is demonstrated to be rare in the bulge \citep{2007A&A...463..251U,2015MNRAS.451.1750U}, and especially less common among the short-period Miras we use in our analysis. The effect of C-rich	Miras is a point of interest, and merits further investigation, especially in the Bulge; however, here we assume that removing long-period Miras nullifies the effect of C-rich contamination.
	
	To summarize, we make the following choices in choosing the best distance estimate for the Galactic Centre:
	\begin{enumerate}[leftmargin=!,labelindent=15pt,itemindent=-5pt]
		\item Only Miras from the SAAO and MACHO surveys are used, since the OGLE catalogue may be affected by saturation.
		\item We measure reddening using the \cite{2000MNRAS.319..728W} period-color relation, since it measures total extinction and is more consistent with the interstellar dust maps at low periods.
		\item The reddening estimate is paired with the \cite{2009ApJ...696.1407N} extinction to reddening ratio.
		\item Only stars with $\log P < 2.6$ are used for the line fitting, since longer period stars are affected by greater circumstellar extinction and contamination by C-rich Miras.
		\item Referring to Table \ref{tab:dist_ext_nogle}, the above choices give a distance of $R_0 = 8.1$ kpc.
		\item To account for the geometric effect of the Mira latitude distribution, we subtract 0.2 kpc, yielding $R_0 = 7.9$ kpc.
	\end{enumerate}
	
	The statistical uncertainties from the different extinction estimates and fitting the distance modulus total to about $0.2$ kpc. An additional error of 0.2 kpc from zero-point uncertainties is discussed in Section \ref{sec:Zeropoints} below. We exclude the systematic uncertainty associated with using LMC relations, as this error is addressed in Section \ref{sec:Models}. Thus, our best and final distance estimate is $R_0 = 7.9 \pm 0.3$ kpc. This is in good agreement with the measurement from the orbit of the star S2 around Sgr A* of  $R_0 = 8.122 \pm 0.031$ kpc  \citep{2018A&A...615L..15G}, as well as that of the best estimate from other studies, $R_0 = 8.2 \pm 0.1$ kpc \citep{2016ARA&A..54..529B}. 
	
	\subsection{Distance error due zero-point uncertainties}\label{sec:Zeropoints} 
	
	Equation \ref{eqn:whitelock_pl} is derived from a fit to the period-luminosity relation of Miras in the Large Magellanic Cloud, and thus includes two separate sources of error. The original value of the intercept, $\delta$, from \citet{2008MNRAS.386..313W}, is  $(11.241 + \mu_{\rm{LMC}}) \pm 0.026$. The LMC distance estimate from \citet{2013Natur.495...76P} is $18.49 \pm 0.05$. The sum of these two in quadrature is $\sim 0.056$ magnitudes, or $\sim 200$ pc.
	
	This is a zero-point error that is separate and independent from our other sources of uncertainty. Future investigations should be able to adjust the final results accordingly, as the data on the LMC eventually improves.

	\begin{table}
		\centering
		\caption{Estimates of $R_0$ for different intervals of period and Galactic latitude and longitude and using the \protect\cite{2000MNRAS.319..728W} period-color relation with the \protect\cite{1999hia..book.....G} extinction coefficient. $JHK$ values in the SAAO system were used for the calculations. The sample fitted includes OGLE Miras, excludes the most recent 192 Miras found in the MACHO survey, and does not make a cut on Miras outside a spherical volume of 4 kpc around the Galactic centre. These values are all in close agreement with the estimates derived by \citet{2016MNRAS.455.2216C}.}
		\renewcommand{\arraystretch}{1.2}
		\begin{tabular}{l l l l l}
			\hline
			& \multicolumn{2}{l}{\textbf{All longitudes}} & \multicolumn{2}{l}{$\boldsymbol{|l| \leq 3\degree}$} \\
			\textbf{Period Range} & $R_0$ (kpc) & N & $R_0$ (kpc) & N \\
			\hline
			& \multicolumn{4}{l}{$\boldsymbol{|b| < 4.5 \degree}$} \\
			2.1 < $\log P$ < 2.2 & 8.857 & 113 & 8.664 & 67 \\
			2.2 < $\log P$ < 2.3 & 9.015 & 310 & 8.689 & 198 \\
			2.1 < $\log P$ < 2.3 & 8.969 & 423 & 8.683 & 265 \\
			2.3 < $\log P$ < 2.4 & 8.879 & 704 & 8.744 & 437 \\
			2.4 < $\log P$ < 2.5 & 9.144 & 1415 & 8.906 & 814 \\
			2.5 < $\log P$ < 2.6 & 8.820 & 1785 & 8.721 & 1021 \\
			2.1 < $\log P$ < 2.6 & 8.949 & 4327 & 8.778 & 2537 \\
			2.6 < $\log P$ < 2.7 & 8.368 & 1104 & 8.325 & 622 \\
			& \multicolumn{4}{l}{$\boldsymbol{|b| > 4.5 \degree}$} \\
			2.1 < $\log P$ < 2.6 & 8.734 & 1163 & 9.004 & 293 \\
			2.6 < $\log P$ < 2.7 & 7.985 & 275 & 8.444 & 91 \\
			& \multicolumn{4}{l}{$\boldsymbol{|b| > 6.0 \degree}$} \\
			2.1 < $\log P$ < 2.6 & 8.233 & 468 & 8.397 & 44 \\
			2.6 < $\log P$ < 2.7 & 7.514 & 89 & 8.712 & 19 \\
			& \multicolumn{4}{l}{\textbf{SAAO Miras only}} \\
			2.1 < $\log P$ < 2.6 & 8.150 & 512 &  & \\
			2.6 < $\log P$ < 2.7 & 7.389 & 87 &  & \\
			\hline
		\end{tabular}
		\label{tab:dist_reprod}
	\end{table}

\begin{table}
		\centering
		\caption{$R_0$ estimates for each catalogue separately. As expected, the OGLE catalogue gives a significantly higher distance estimate compared to SAAO and MACHO.}
		\renewcommand{\arraystretch}{1.2}
		\begin{tabular}{l l l l l}
			\hline
			& \multicolumn{2}{l}{$\boldsymbol{2.1 < \log P < 2.6}$} & \multicolumn{2}{l}{$\boldsymbol{2.6 < \log P < 2.7}$} \\
			\textbf{Catalogue} & $R_0$ (kpc) & N & $R_0$ (kpc) & N \\
			\hline
			SAAO & 7.854 & 410 & 7.479 & 71 \\
			OGLE & 8.580 & 3581 & 8.083 & 977 \\
			MACHO & 8.252 & 834 & 7.589 & 201 \\
			\hline
		\end{tabular}
		\label{tab:dist_compare}
	\end{table}
	
	\begin{table}{}
		\centering
		\caption{Estimates of $R_0$ using different extinction maps and excluding OGLE survey Miras. Names in parentheses indicate the coefficient used to convert between reddening and extinction.}
		\renewcommand{\arraystretch}{1.2}
		\begin{tabular}{l l l l l}
			\hline
			& \multicolumn{2}{l}{$\boldsymbol{2.1 < \log P < 2.6}$} & \multicolumn{2}{l}{$\boldsymbol{2.6 < \log P < 2.7}$} \\
			\textbf{Method} & $R_0$ (kpc) & N & $R_0$ (kpc) & N \\
			\hline
			Whitelock (Nishiyama) & 8.098 & 1245 & 7.538 & 272 \\
			Glass (Nishiyama) & 7.969 & 1255 & 7.438 & 275 \\
			Whitelock (Glass) & 8.049 & 1244 & 7.480 & 272 \\
			Glass (Glass) & 7.927 & 1257 & 7.395 & 273 \\
			Gonzalez & 8.725 & 600 & 8.224 & 134 \\
			Schlafly & 8.151 & 1250 & 7.998 & 263 \\
			\hline
		\end{tabular}
		\label{tab:dist_ext_nogle}
	\end{table}
	
	\begin{table}{}
		\centering
		\caption{Similar to Table \ref{tab:dist_ext_nogle}, but including OGLE catalogue stars. The $R_0$ values of Table \ref{tab:dist_ext_nogle} agree more closely with estimates reported by other studies than the values shown here, possibly because of saturation in the OGLE catalogue.}
		\renewcommand{\arraystretch}{1.2}
		\begin{tabular}{l l l l l}
			\hline
			& \multicolumn{2}{l}{$\boldsymbol{2.1 < \log P < 2.6}$} & \multicolumn{2}{l}{$\boldsymbol{2.6 < \log P < 2.7}$} \\
			\textbf{Method} & $R_0$ (kpc) & N & $R_0$ (kpc) & N \\
			\hline
			Whitelock (Nishiyama) & 8.618 & 3999 & 8.150 & 1012\\
			Glass (Nishiyama) & 8.474 & 4045 & 8.006 & 1035\\
			Whitelock (Glass) & 8.542 & 4020 & 8.052 & 1028\\
			Glass (Glass) & 8.387 & 4066 & 7.883 & 1046\\
			Gonzalez & 8.854 & 3838 & 8.796 & 851 \\
			Schlafly & 8.719 & 3947 & 8.712 & 874 \\
			\hline
		\end{tabular}
		\label{tab:dist_ext}
	\end{table}

\section{Effects of age and metallicity}
	\label{sec:Models}
	
	In order to investigate the dependence of the Mira $K_s$--$\log P$ relations on age, metallicity, and helium abundance, we have derived theoretical $K_s$--$\log P$ relations using the linear, non-adiabatic, radial pulsation code described in \cite{2014MNRAS.440.2576W}. This code has been used to identify the pulsation modes associated with the five most prominent period-luminosity sequences exhibited by pulsating AGB stars in the LMC \citep{2015MNRAS.448.3829W, 2017ApJ...847..139T}. It is assumed that the Mira variables are radial fundamental mode pulsators \citep{2015MNRAS.448.3829W}.  We expect that differential effects on the period caused by changes in mass (age), metallicity and helium abundance are reliable, although the absolute value of the period will have some uncertainty due to the uncertainty in the mixing length parameter of convection.
	
	The age associated with each $K_s$--$\log P$ model relation is calculated using the equation
	\begin{equation}
	    	\log \left( \frac{M}{M_\odot} \right) = 0.026 + 0.126 \,\rm{[Fe/H]} 
	- 0.276 \log \left( \frac{t}{10} \right) - 0.937 (Y - 0.27)
	\end{equation}
	
	from \cite{2012AcA....62...33N}. Here, [Fe/H] is the metallicity, $Y$ is the helium mass fraction, and $t$ is the age in Gyr. Bolometric corrections are used to obtain the $J$ and $K_s$ magnitudes for each track using the relation
	\begin{equation}
	M_{Ks} = M_{\rm{bol},\odot} - 2.5 \log \left( L/L_\odot \right) - BC_{Ks},
	\end{equation}
	where $BC$ stands for bolometric correction. We have used tables of bolometric corrections from \cite{2014MNRAS.444..392C} and fixed $M_{\rm{bol},\odot} = 4.75$. A similar equation is used for $M_J$. While \cite{2014MNRAS.444..392C} did find some discrepancies between predictions and observations of giant stars, these are our best available estimates for the bolometric corrections, so we choose to ignore these considerations.
	
\begin{figure*}
\centering     
\subfigure{\label{fig:PK_y}\includegraphics[width=80mm]{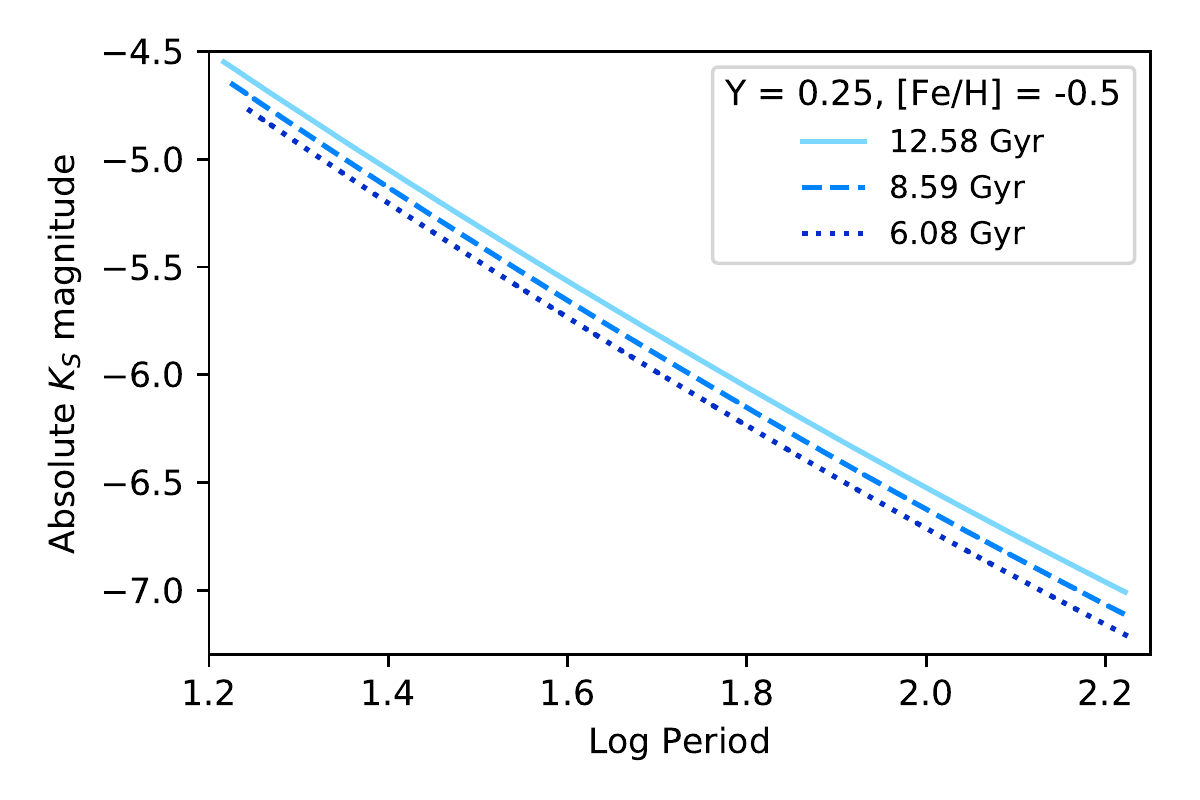}}
\subfigure{\label{fig:PC_y}\includegraphics[width=80mm]{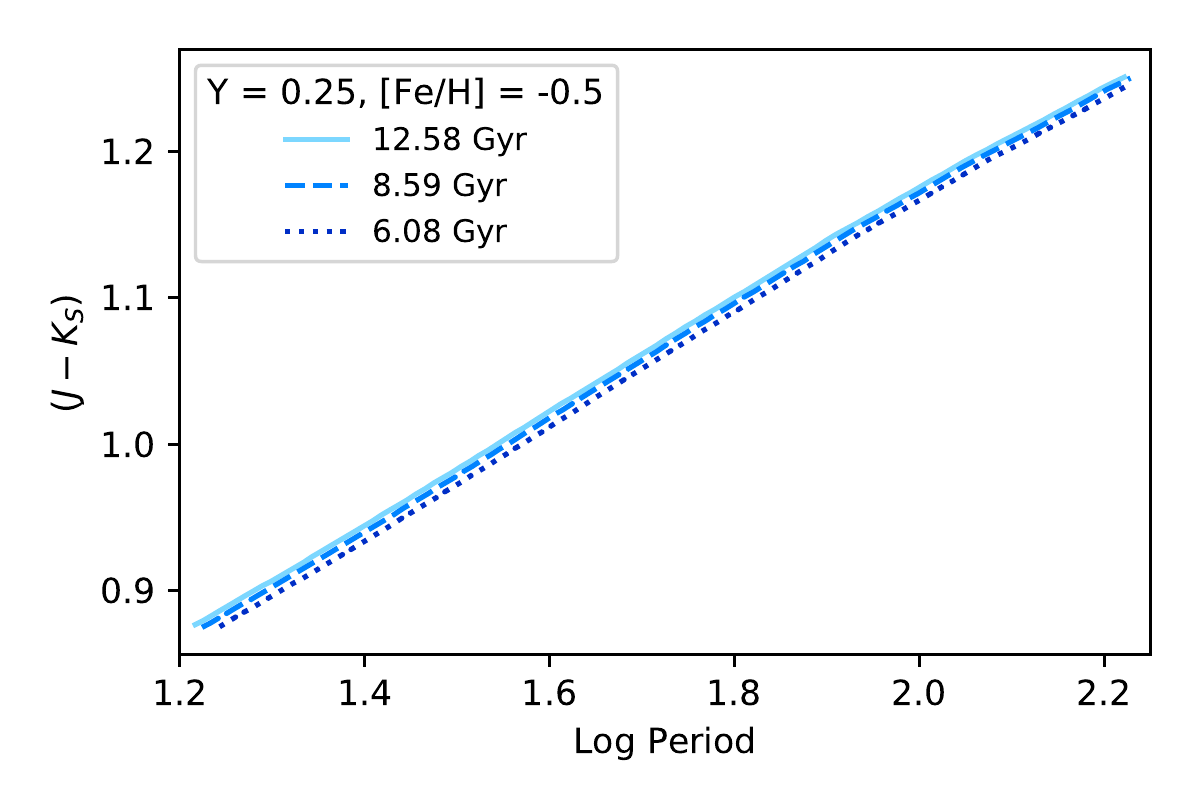}}
\subfigure{\label{fig:PK_age}\includegraphics[width=80mm]{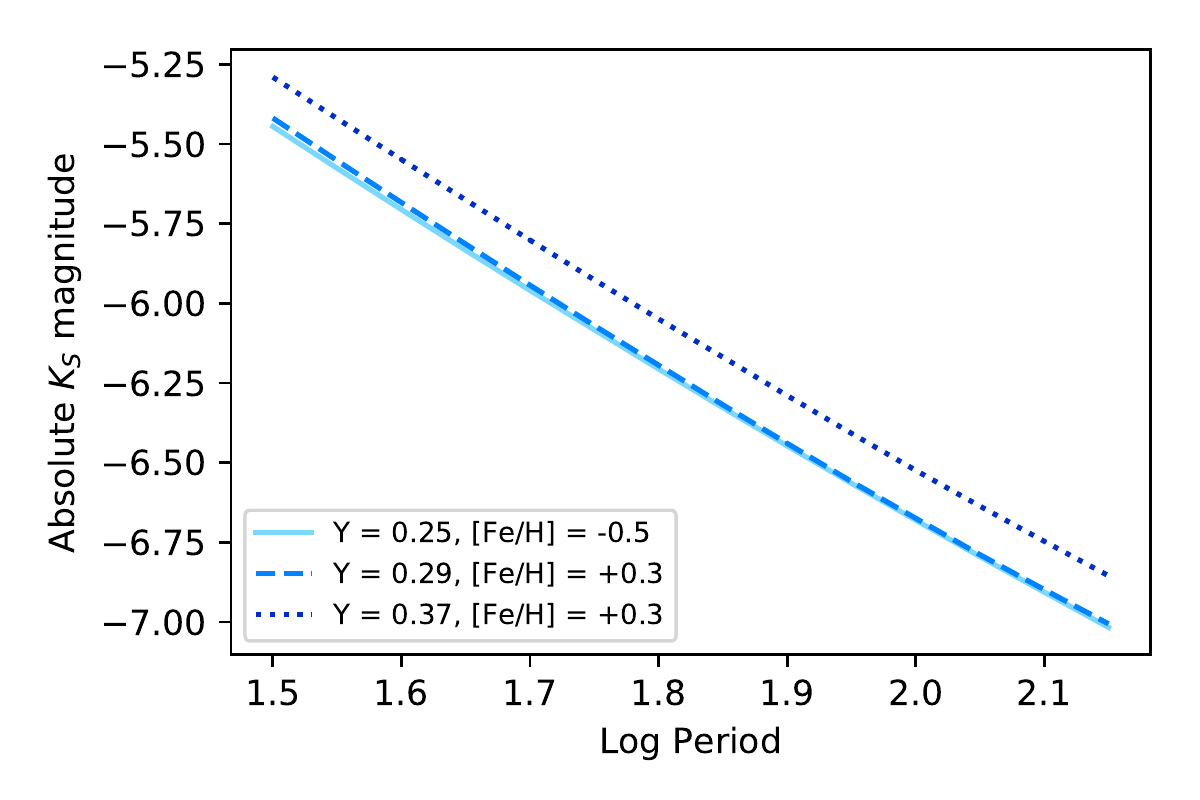}}
\subfigure{\label{fig:PC_age}\includegraphics[width=80mm]{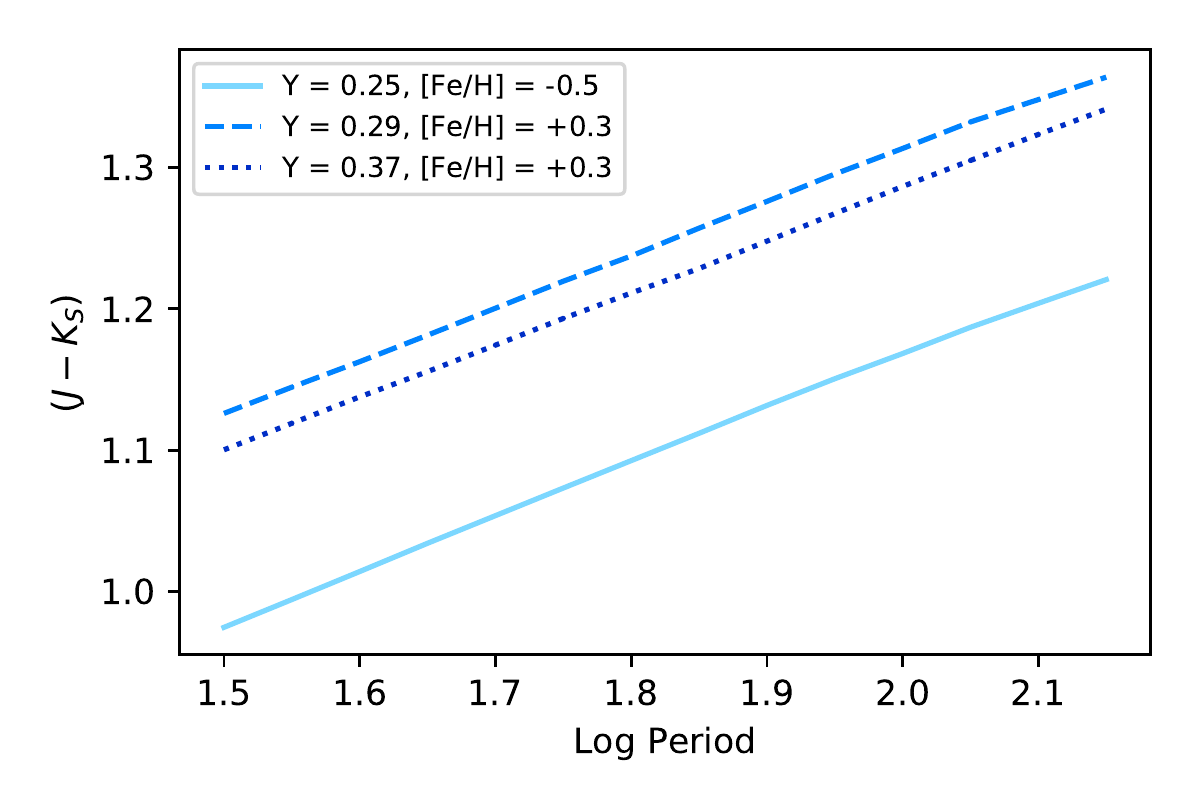}}
\caption{Period-luminosity ($K_s$) and period-color relations obtained from the theoretical AGB models. The left panels show the period-$K_s$ magnitude relations and the right panels show the period-color relations. The top panels show the relations at fixed metallicity and helium, with varying age. The bottom panels show the relations at fixed age, with varying helium and metallicity. While the period-$K_s$ magnitude relation depends on all three quantities, the period-color relation has a weak dependence on age and depends most strongly on metallicity.}
		\label{fig:PL_dependence}
\end{figure*}

\begin{figure}
	    \hspace{-5mm}
		\includegraphics[width=0.55\textwidth]{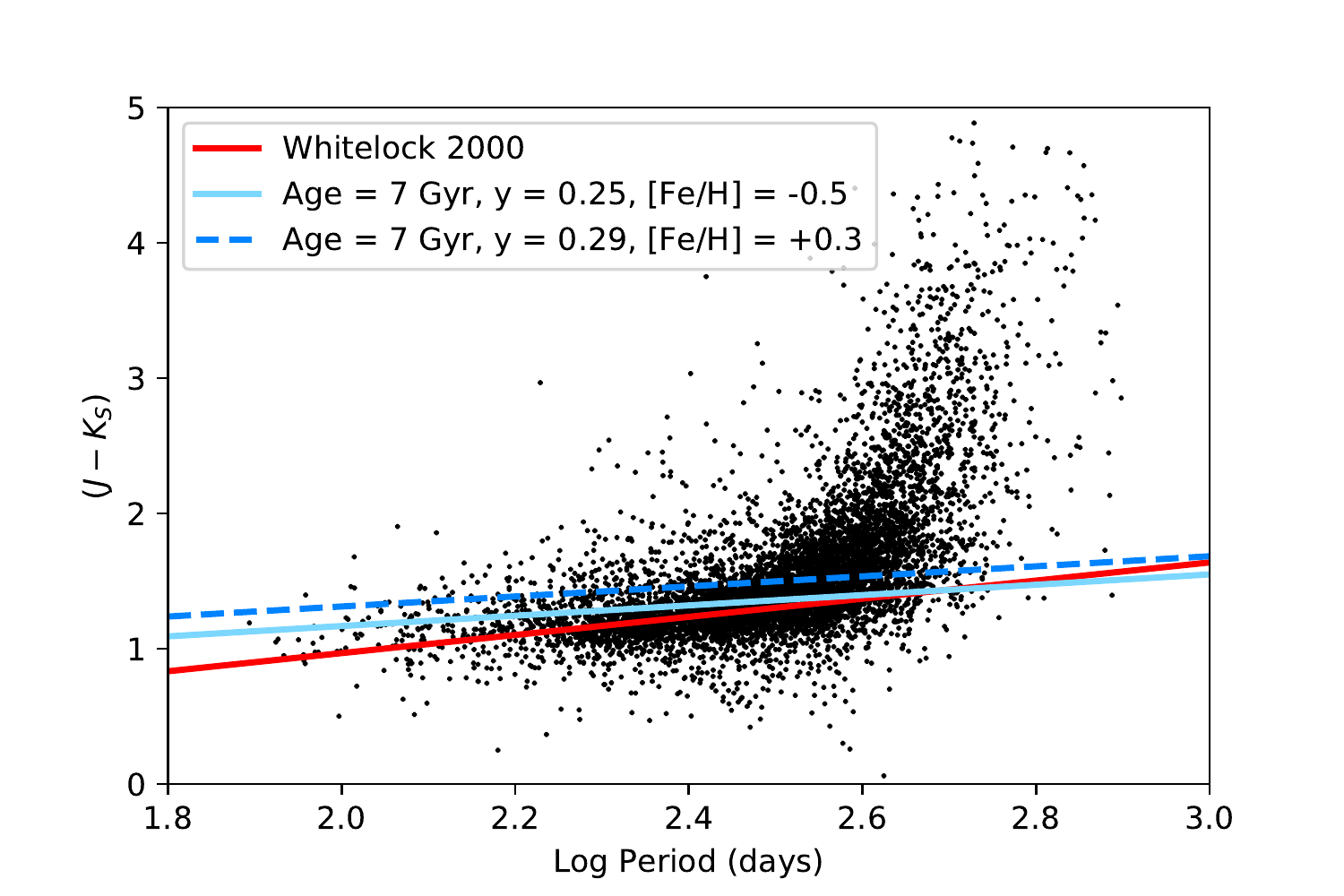}
		\caption{Predicted period-color relations for the computational models. We show two tracks of different chemical composition in blue and compare them to the \protect\cite{2000MNRAS.319..728W} relation in red. The predicted period-color relations nearly match the slope of the empirical relation, but are offset in the zero-point. The data shown in black is the same data we use in Figure \ref{fig:PC_schl+wg}, which has been dereddened using the \protect\cite{2011ApJ...737..103S} map.}
		\label{fig:PC_compare}
	\end{figure}

	Using the models, which give $P$ for input values of $L$, $M$, $Y$ and [Fe/H], and these relations, we can determine the theoretical period-luminosity and period-color relations. We fit lines to these relations using a least-squares fit and find that all the tracks are nearly parallel. We then examine how the period-luminosity and period-color relations of the theoretical tracks are affected by changes in age, metallicity, or helium abundance. As shown in Figure \ref{fig:PL_dependence}, the period-$K_s$ magnitude relation depends on all three quantities, while the period-color relation depends rather weakly on age and most strongly on metallicity. Figure \ref{fig:PC_compare} shows that the theoretical period-color relations are comparable to the \cite{2000MNRAS.319..728W} relation we use in our analysis. The offsets are small, but real, so we use the theoretical models only for predictions relative to the LMC, as such relative predictions are less sensitive to zero-point issues. In addition, the offset size depends on chemical composition. This motivates future research to measure the metallicities of Bulge Miras either directly or to infer them from kinematics. If the Miras trace the metal-rich bulge population, they will have a lower radial velocity dispersion than if they trace the metal-poor bulge population \citep{2013MNRAS.432.2092N,2014A&A...563A..15B}.

	\begin{table}
		\centering
		\caption{Estimated extinction and $K_s$ magnitude errors, using the LMC track ($Y=0.25$, $\rm{[Fe/H]}=-0.5$, $\rm{Age}=6$ Gyr) as the calibrator. The errors are calculated at the fixed value $\log P = 2.4$. Distance offsets are calculated assuming a distance modulus of 14.6, which is about 8.3 kpc. Negative values of the offsets correspond to the stars being intrinsically bluer and fainter.}
		\renewcommand{\arraystretch}{1.2}
		\begin{tabular}{l|l l|l l}
			\hline
			\textbf{Y, [Fe/H], Age (Gyr)} & $A_{err}$ & Offset (kpc) & $M_{Ks,err}$ & Offset (kpc) \\
			\hline
			0.25, -0.5, 7 & -0.001 & 0.003 & 0.036 & -0.137 \\
			0.29, 0.3, 7 & -0.083 & 0.323 & 0.032 & -0.122 \\
			0.37, 0.3, 7 & -0.068 & 0.263 & 0.197 & -0.721 \\
			0.25, -0.5, 10 & -0.003 & 0.012 & 0.137 & -0.509 \\
			0.29, 0.3, 10 & -0.087 & 0.340 & 0.132 & -0.491 \\
			0.37, 0.3, 10 & -0.070 & 0.272 & 0.294 & -1.054 \\
			\hline
			0.27, -0.1, 10 (MW) & -0.045 & 0.174 & 0.135 & -0.500 \\
			\hline
		\end{tabular}
		\label{tab:dist_off}
	\end{table}
	
		From these models, we deduce what the offsets in $R_0$ would be if we assumed different values for age and metallicity, relative to the $R_0$ that would be calculated using LMC relations. The extended star formation history in the LMC means that the period-luminosity relation there is defined by multiple masses and ages, as shown by \citet{1996MNRAS.282..958W} and \citet{2015MNRAS.448.3829W}. However, since the predicted near-infrared colors and magnitudes have an almost linear dependence on age, helium, and metallicity, we conclude that the color and magnitude at the mean age and metallicity are equal to the mean color and magnitude of the population as a whole. Using the models in this way also assumes that the Miras trace the bulk stellar population of a galaxy. We can see evidence for this assumption if we examine Figure 2, panel D, of \cite{2000ASSL..255..229F}, which shows the derived metallicity distribution of Bulge Miras assuming a period-color relation calibrated off of Galactic globular clusters. This distribution is negative skewed and centered at a slightly sub-solar metallicity, spanning the range $-1.50 \leq \rm{[Fe/H]} \leq +0.50$. Comparing this to Figure 6 of \cite{2013MNRAS.430..836N}, which is the Bulge metallicity distribution function derived from spectroscopic metallicities of Bulge red clump stars, we can see that the two plots are visually quite similar. This suggests that the work of \cite{2000ASSL..255..229F} holds up two decades later, and that the Miras approximately trace the bulk population.
	
	We take the theoretical track at age 6 Gyr, $0.25$ helium abundance, and $\rm{[Fe/H]} = -0.5$ to represent the LMC \citep{2009AJ....138.1243H, 2013MNRAS.431..364W,2016MNRAS.455.1855C}. Using the line fits, we first calculate what the discrepancies in extinction and $M_{Ks}$ would be between the LMC track and other tracks at fixed period (See Table \ref{tab:dist_off}). Since the tracks are nearly parallel, these discrepancies remain about the same across different periods. We then convert these errors to distance offsets using $\log(d) = 1 + \mu/5$, where $d$ is in pc and we assume $\mu = 14.6$. Lastly, we interpolate the errors and offsets of the Milky Way relations relative to the LMC, where the track with age 10 Gyr, $0.27$ helium, and $\rm{[Fe/H]} = -0.1$ is used to approximate the Milky Way Bulge \citep{2016PASA...33...23N, 2017A&A...605A..89B}. The results of \cite{2016MNRAS.455.1855C} and \cite{2017A&A...605A..89B} indicate that the Bulge is at least $\sim 0.30$ dex more metal-rich, on average, than the LMC, and the difference may be as much as 0.40 dex once the alpha-enhancements are taken into account. 
	
	We find that the Milky Way track underestimates extinction by $-0.045$, which would cause our previously calculated $R_0$ to increase by about $0.2$ kpc. On the other hand, the Milky Way track also overestimates the $K_s$ magnitude by $0.135$, which would cause the distance estimate to decrease by $\sim 0.5$ kpc. Overall, an $R_0$ value calculated using Milky Way quantities should be smaller than the $R_0$ calculated assuming LMC relations by $\sim 0.3$ kpc.
	
	While age is a less important factor than metallicity, we would like to make age corrections as well. We assume that the age of the Bulge is twice the age of the LMC, though the exact age distribution of the bulge is still controversial. For example, \cite{2011ApJ...735...37C} rule out a population younger than 5 Gyr, whereas \cite{2017A&A...605A..89B} estimate that $\sim 20\%$ of bulge stars are younger than 5 Gyr. As we have already emphasized, in the regime where the corrections are linear with age, it is the mean value of the age that matters most, and that value is taken to be 10 Gyr, as opposed to LMC's 6 Gyr.
	
	Finally, we determine the coefficients relating changes in $J$ and $K_s$ magnitudes to age, metallicity, and helium abundance. We also include equations for changes in bolometric magnitude, effective temperature, $V$, $I$, and $H$ magnitudes, since space-based studies of Miras will make use of these quantities (the $V$, $I$, and $H$ magnitudes were derived from bolometric corrections in the same way as $J$ and $K_s$). At a fixed period of $\log P = 2.4$, the changes in these quantities can be calculated using Table \ref{tab:diff}. The covariances between changes in age, metallicity, and helium abundance are small. In addition, since the period-luminosity and period-color relations in the $J$, $H$, and $K_s$ bands are parallel for different values of age, metallicity, and helium abundance, the coefficients in Table (\ref{tab:diff}) vary by a negligible amount at different values of $\log P$, so the relations for infrared bands can be used for Miras over a range of $\log P$. However, the $M_{\rm{bol}}$, $\log T_{\rm{eff}}$, $V$, and $I$ equations show a much greater variability. For example, at $\log P = 2.0$, the coefficient relating $\Delta M_V$ and [Fe/H] is 4.297, while at $\log P = 2.6$ the coefficient is 7.314. Thus, this variable dependence of the $M_{\rm{bol}}$, $\log T_{\rm{eff}}$, $V$, and $I$ equations on metallicity makes it difficult to apply them across a range of periods.
	\begin{table}
		\centering
		\caption{Coefficients relating changes in various magnitudes to changes in metallicity, age, and helium at period of $\log P = 2.4$. For example, according to this table, if the age of a star changes by 1 Gyr, then its $J$ magnitude will change by 0.034 mag. The infrared colors can be used for Miras over a large range of periods; however, the $M_{\rm{bol}}$, $\log T_{\rm{eff}}$, $M_V$, and $M_I$ equations show greater variability across periods, so their coefficients are less generally useful.}
		\begin{tabular}{l l l l}
			\hline
		    $\Delta M$ & $\Delta M / \Delta \rm{ \left[ Fe/H \right]}$ & $\Delta M / \frac{\Delta t}{\rm{Gyr}}$ & $\Delta M / \frac{\Delta Y}{0.01}$ \\
			\hline
			$\Delta M_{\rm{bol}}$ & $-7.974 \pm 1.084$ & $0.398 \pm 1.198$ & $-0.028$ \\
			$\Delta \log T_{\rm{eff}}$ & $5.337 \pm 11.091$ & $1.230 \pm 2.422$ & $-0.233$ \\
			$\Delta M_V$ & $6.251 \pm 0.057$ & $0.064 \pm 0.013$ & $-0.044$ \\
			$\Delta M_I$ & $2.606 \pm 0.054$ & $0.052 \pm 0.012$ & $-0.022$ \\
			$\Delta M_J$ & $0.085 \pm 0.002$ & $0.035 \pm 0.001$ & $0.017$ \\
			$\Delta M_H$ & $0.024 \pm 0.001$ & $0.033 \pm 0.001$ & $0.020$ \\
			$\Delta M_{Ks}$ & $-0.109 \pm 0.001$ & $0.033 \pm 0.001$ & $0.021$ \\
			\hline
		\end{tabular}
		\label{tab:diff}
	\end{table}
	
	The predicted metallicity dependence in $K_{s}$ is $\sim 4 \times$ smaller than that reported by \citet{2005A&A...443..143G} in their comparison of SMC, LMC, and Bulge data. That may be due to the effect of saturation biasing the size and distribution of the OGLE-II sample. Indeed, they report a distance to the Galactic centre of 9.0 kpc under the assumption of $\mu_{LMC}=18.50$, which is now ruled out by Galactic structure studies and well-explained by the issue of saturation. 
	
		\subsection{Comparison to solar neighbourhood Miras}
	\label{sec:solar neighbourhood}
	
\begin{figure*}
		\centering
		\includegraphics[width=0.9\textwidth]{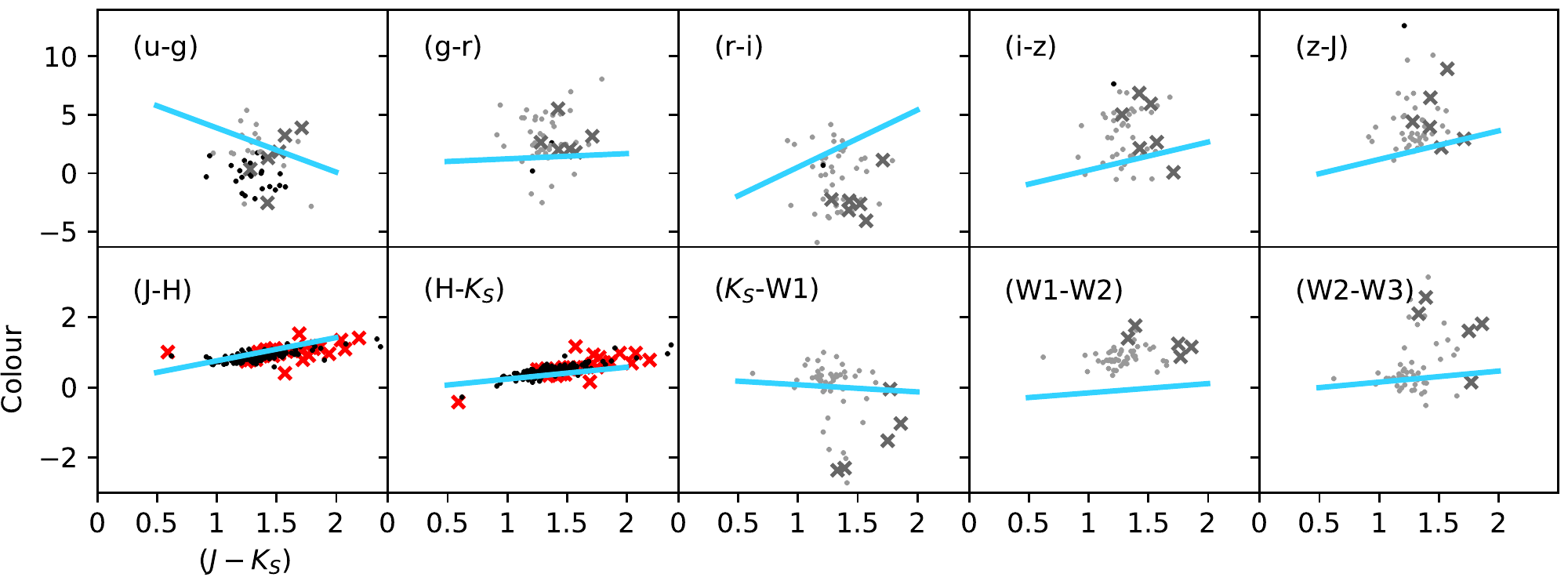}
		\caption{Color-color diagrams of solar neighbourhood Miras, where dots represent Miras with $\log P < 2.6$ and Xs represent Miras with $\log P > 2.6$. Grey dots and Xs indicate measurements that may be affected by saturation, while black dots and red Xs are unaffected, and the lines indicate the colors predicted by the AGB models for the Milky Way ($Y=0.27$, $\rm{[Fe/H]}=-0.1$, $\rm{Age}=10$ Gyr). The y-axis colors are given by the text in each plot, while the x-axis color is always $(J-K_s)$. The theoretical tracks seem to model the data fairly well in near-infrared colors, but the diagrams showing $ugriz$ and WISE colors indicate that there are still significant discrepancies. This could either be due to the uncertainty in the data due to saturation or the model's failure to reproduce certain colors. Therefore, calculations based on these models should be considered only approximate.}
		\label{fig:cc_diagrams}
	\end{figure*}

	Figure \ref{fig:cc_diagrams} shows color-color diagrams for solar-neighbourhood Miras, with lines indicating the colors predicted by the track used to approximate the metallicity and age of Milky Way Miras. In this plot, we have included colors from WISE and SDSS in order to explore the uses and limitations of this extra photometric information. The theoretical bolometric corrections used to calculate the various colors are computed in the same way as the $J$ and $K_s$ bolometric corrections described above. In addition, since these stars are within the solar neighbourhood, several of the WISE and SDSS magnitudes suffer from saturation, although WISE profile-fitting photometry is still able to extract reliable estimates for most of these measurements \citep{2012wise.rept....1C}. Sources affected by saturation are denoted in grey and blue colors (\citealt{2014MNRAS.442.3361N} discuss saturation of WISE sources in great detail).  While the theoretical colors well match the data in the near-infrared, the tracks do not match nearly as well in the optical or mid-infrared. This discrepancy is either due to the uncertainties in the colors caused by saturation issues, or a deficiency in the model. If it is the latter, then it indicates that the dependencies we get for the period-luminosity relations only give relative changes, and the coefficients in Table \ref{tab:diff}  should be considered uncertain. Table \ref{tab:diff} gives approximate values, since they are based on AGB models and not specifically on Miras. 
	
	The mean value of $(H-W2)$ for the solar-neighbourhood Miras is 1.29 magnitudes. Referring back to Section \ref{sec:RJCE}, this is considerably higher than the 0.08 magnitude mean assumed by the Rayleigh-Jeans color excess method of calculating extinction. Therefore, the colors of the Miras we use give overly large extinction estimates, which explains why the median star distance was only 5.22 kpc, justifying our decision to not include the Rayleigh-Jeans method in our analysis.
	
	The predicted colors from \cite{2014MNRAS.444..392C} should not be expected to work for the C-rich Miras, given that these have significant features unaccounted for by those models. Long-period Miras (which are more likely to be C-rich) are denoted by X's in Figure \ref{fig:cc_diagrams}. They are indeed less well-fit by the predicted relations, though this discrepancy can only explain some of the observed scatter.

	\subsection{Discussion of the predicted distance offset}
	\label{sec:offsetprediction}
	The predicted distance offset for Galactic bulge Miras of $\sim$0.30 kpc would result in a final inferred distance of about $7.6 \pm 0.3$ kpc, which is mildly in tension with the literature estimate of $8.2 \pm 0.1$ kpc, at the 2.0-$\sigma$ level \citep{2016ARA&A..54..529B}. Its tension with the recent measurement from the orbit of the star S2, $8.122 \pm 0.031$ kpc \citep{2018A&A...615L..15G}, is a milder 1.73-$\sigma$. We discuss several reasons why that might be, and how the issue might be resolved.
	
	The first is that the offset is still small enough that one can reasonably suspect it to be a fluke. The respective odds of a 2.0 and 1.73-$\sigma$ event are  $\sim 4.5$\% and $\sim 8.4\%$ if one assumes Gaussian errors, which might not apply here. 
	
	A second possibility is that the spatial heterogeneity of the Mira sample could also provide its own errors, for example if the sample happens to correlate with reddening law variations, which are themselves heterogeneous with direction (e.g. \citealt{2016ApJ...821...78S,2017ApJ...849L..13A}). 
	
	A third possibility is that the Bulge Mira population's sampling of stellar population parameters is a function of period, at least for short-period Miras ($\log P \leq 2.6$). For example, \citet{2016MNRAS.455.2216C} showed in their Section 6 that the spatial distribution of shorter-period Miras is consistent with hotter kinematics, which are associated with the more metal-poor and thus likely older component of the bulge \citep{2012ApJ...750..169P,2018arXiv180401103C}. This and other evidence suggests that a star with a particular initial mass, metallicity, and initial helium abundance becomes a Mira whose period evolves little over time.
	
	These concerns, as well as the model predictions which they address, are testable. \citet{2018ApJ...855L...9D} have accurately and precisely measured the abundance ([Fe/H] $= -0.55 \pm 0.15$) of a 202-day period Mira in the globular cluster NGC 5927 by means of a brief, 300 second spectroscopic exposure. The Mira is a little fainter ($K_{s}=8.9 \pm 0.15$) than most Bulge Miras, and thus the methodology of \citet{2018ApJ...855L...9D} can be applied to Bulge Miras. It is thus demonstrably possible to ascertain these predictions of asymptotic giant branch models, and whether or not subcomponents of the stellar population with specific ages and metallicities can produce Miras with a range of periods. 
	
	We note that \citet{2018ApJ...855L...9D} also measured [Na/Fe] in this Mira, though they state that further investigation is needed to confirm if this specific abundance is reliable. Should these calibrations be completed, there will be a means to infer helium abundances for globular cluster Miras, as helium and sodium abundance offsets correlate on the asymptotic giant branches of globular clusters \citep{2017ApJ...843...66M}. This will provide a means to not just test the dependence on age and metallicity discussed above, but also that on helium abundance.

	\section{Conclusion}
	\label{sec:Conclusion}
	
	In this study, we examine the validity of making distance measurements using Mira variables by using them to measure the distance to the Galactic centre, as well as probing the dependence of the Mira period-luminosity relation on a galaxy's age and composition. In selecting an ideal sample of Bulge Miras for fitting the distance to the Galactic centre, we find that the OGLE catalogue \citep{2013AcA....63...21S} has low completeness for brighter stars (i.e., stars on the near side of the Bulge) due to saturation, making it unsuitable to use for our distance study. 
	
	In comparing several methods of estimating extinction, we find that color-based techniques for calculating extinction towards Miras work better than Galactic dust maps. That may be because former method is less sensitive to the effects of circumstellar extinction. After applying such a method, choosing stars with periods $\log P < 2.6$, and making a geometric correction, we determine that our best estimate for the distance to the Galactic centre is $R_0 = 7.9 \pm 0.3$ kpc, which is in good agreement with measurements of $R_0$ based on other methods in the literature \citep{2016ARA&A..54..529B,2018A&A...615L..15G}. 
	
	We use theoretical tracks and bolometric corrections to model Mira period-luminosity and period-color relations and study their dependence on age and chemical composition. In comparing the colors predicted by these models to the colors of solar-neighborhood Miras, we find discrepancies in the optical and near-infrared photometric bands, which is either due to saturation or deficiencies in the models. This suggests that the relations we derive should only be used as approximations.
	
	However, assuming that these models are valid for Galactic Miras, we find that there is a non-negligible dependence of the relations on metallicity and helium, with a smaller effect from stellar age. Since the Milky Way Bulge is about twice as old and twice as metal-rich as the LMC, using relations based on the LMC should cause an overestimate of $R_0$ on the order of $\sim 0.3$ kpc. This has not been validated by our analysis, and we look forward to more precise tests from future investigations. Thus, as we strive to use Mira variables to make increasingly precise distance estimates, both within and outside of the Galaxy, accurately determining the variation of the period-luminosity relations from galaxy to galaxy will become more important.

	\section*{Acknowledgments}
	
	WQ was supported by NASA's Maryland Space Grant Consortium.
	DMN was supported by the Allan C. and Dorothy H. Davis Fellowship. 
	NLZ was supported by the Johns Hopkins University Catalyst Award.
	LC gratefully acknowledges support from the Australian Research Council (grants DP150100250, FT160100402).
	We would like to thank Christopher Wegg for helpful discussions of systematics in bulge distance estimates.
	We would like to thank the referee for constructive feedback on the manuscript. 
	We made use of the VizieR and SIMBAD databases of the CDS in preparing this paper. 
	This paper uses observations made at the South African Astronomical Observatory (SAAO). 
	The OGLE project has received funding from the National Science Centre, Poland, grant MAESTRO 2014/14/A/S3T9/00121 to AU. 
	This paper utilizes public domain data obtained by the MACHO Project, jointly funded by the US Department of Energy through the University of California, Lawrence Livermore National Laboratory under contract No. W-7405-Eng-48, by the National Science Foundation through the Center for Particle Astrophysics of the University of California under cooperative agreement AST-8809616, and by the Mount Stromlo and Siding Spring Observatory, part of the Australian National University. 
	This publication makes use of data products from the Two Micron All Sky Survey, which is a joint project of the University of Massachusetts and the Infrared Processing and Analysis Center/California Institute of Technology, funded by the National Aeronautics and Space Administration and the National Science Foundation. 
	This publication makes use of data products from the Wide-field Infrared Survey Explorer, which is a joint project of the University of California, Los Angeles, and the Jet Propulsion Laboratory/California Institute of Technology, funded by the National Aeronautics and Space Administration.
	

	\label{lastpage}
	
\end{document}